# Python vs. R: A Text Mining Approach for analyzing the Research Trends in Scopus Database

Neeraj Bhanot[1]*, Harwinder Singh[2], Divyansu Sharma[3], Harshit Jain[4] and Shreyansh Jain[4]

[1]Assistant Professor, Quantitative Methods and Operations Management, Indian Institute of Management Amritsar, Amritsar – 143105

[2]Professor, Department of Mechanical Engineering, Guru Nanak Dev Engineering College, Ludhiana-141006

[3]UG Student, Department of Mechanical Engineering, Dr. B R Ambedkar National Institute of Technology Jalandhar, Jalandhar – 144011

[4]UG Student, Department of Industrial and Production Engineering, Dr. B R Ambedkar National Institute of Technology Jalandhar, Jalandhar – 144011

*Corresponding Author: neeraj.bhanot@iimamritsar.ac.in

**Abstract**: In the contemporary world, with the incubation of advanced technologies and tremendous outbursts of research works, analyzing big data to incorporate research strategies becomes more helpful using the tools and techniques presented in the current research scenario. This paper indeed tries to tackle the most prominent challenges relating to big data analysis by utilizing a text mining approach to analyze research data published in the field of production management as a case to begin with. The study has been conducted by considering research data of International Journal of Production Research (IJPR) indexed in Scopus between 1961-2017 by dividing the analysis incurred into 3 fragments being 1961-1990, 1991-2010 and finally 2011-2017 as a case to highlight the focus of journal. This has indeed provided multi-faceted benefits such as increasing the effectiveness of the procured data with well-established comparisons between R and Python Programming along with providing detailed research trends on the research work incubated. The results of the study highlighted some most prominent topics in the existing IJPR literature such as "system's optimization, supplier selection, process design", etc. providing well-established details relating to ongoing research works. The study also compared both languages suiting to a particular field of study for better comprehension and vastness of the research topics. The current research work is one of the part of a copyright work with registration number "SW-10310/2018" titled "Program for Analyzing Key Trends in Research Data-set". It has been designed in Python for carrying out detailed content analysis based on the available research database in bib format as in the current context it has been applied for IJPR journal and can be replicated on articles of any domain found using keyword search.

## 1. Introduction

Data handling in the contemporized world is one of those issues which have been confronted by lots organizations and research inventories etc. since the task of managing the loads of information turns out to be a very cumbersome job. However, in the present scenario data analyzing is not at all a herculean task, thanks to the devised tools and techniques. In the contemporary state, with the incorporation of text mining approach that aggregates and structures data in a largely automated manner, analyzing big data becomes more feasible with an essence of great accuracy and objectivity

(Amado and Cortez, 1996). The primarily focus in this research paper is to provide a well-defined analysis encompassing the 3 tiers namely Volume, Variety and Velocity relating to the field of Big Data Analytics. These 3 dividends indeed enrich the quality of research data available from our further studies and procedures.

The International Journal of Production Research (IJPR), is one of most eloquent, well -established and leading journal reporting "manufacturing, production and operations management" (Chaudhry and Lao, 2007) strategies for researchers and professors in "mechanical engineering, industrial and systems engineering, operations research and management science, and business" (Reynolds and Chung, 2007). It incorporates varied aspects of research outlook initiating researchers to improve their efficiency and effectiveness as per the various productive systems. Ultimately IJPR can be called as the "storehouse of research papers dealing with the vastness and extremes of mechanical engineering and its facets" possessing infinite possibilities for its scintillating outcomes in the coming years (Chun and Bidanda, 2013). The modus operandi behind choosing IJPR only, aids to the very fact of the vastness and qualitative research data presented in the journal by various preeminent researchers across the globe. As a part of 55th Anniversary of the Journal in 2017, this research paper is also an outcome of its Scopus indexed relevant content specific papers as specified in the directives of the Journal. The study cumbered in our research paper will ultimately help researchers working in the same segment by enunciating a crux of all the plethora of research informatics impregnated in the journal over a stretch of 60 years. It will also provide a vision for the forthcoming papers by laying out a theoretical modeling in front of the researchers across the globe.

The beauty of the paper lies in the integration of the IJPR journal with the big research data in collaboration making available the research trends present in the journal. This paper strictly entangles about utilizing a text mining approach to analyze qualitative and quantitative data as per the above Journal. This has indeed provided many pragmatic outcomes following the current research trends in a very subtle and subjective manner, differentiating it from the past field related works. The results out casted from our study will help in collaboration of cumbersome research data via the formation of various clusters and then differentiating the common research trends already provided in other papers with the extensive use of Python and R programming to organize and interpret the data for further research analysis (Zaheer han and Tim Vorley, 2016).

Both Python and R programming, are undoubtedly leading platforms for big data analytics, hence efforts have been made to subsidize our research practices using these two computational programming languages. However, to the other side of it major focus has been laid upon to visualize research-oriented study and its comparable efficiencies on analyzing large packets of fragmented data types. While Python is praised for being a general-purpose language with an easy to understand syntax, R's functionality is developed with statisticians in mind, thereby giving it field specific advantages such as great features for data visualization. Hence, it's quite interesting to comprehend how the two platforms would unfold when compared specifically in different areas.

Initially, the IJPR literature from 1961-2017 was considered for in-depth analysis in a consolidated manner but couldn't be achieved due lack of computational efficiencies of the concurrent systems. This undoubtedly raised the need for dividing the literature available into suitable number of standardized fragments. Once again the evident argumentation behind the 3 year slots were mainly based on well-defined strategies such as data available before 1961, could not be analyzed due to very fact that research data available from the Scopus was from 1961 onwards and not before than that. As a result, our study has mainly been incorporated from 1961 only. Now, came the option of dividing data into equal slots of 20 years each so as to maintain uniformity in the literature reviewed throughout these years. However, to our surprise the growth of literature indeed floriferated exponentially and hence a span of 20 years was not eloquent to study the suburbs in the literature. Finally, a tentative period of 30 years followed by 20 and 10 years were chosen respectively just because it approximated the same number of documents over the respective year of stretches. This has indeed helped us in maintaining consistency in the analyzed data over a span of 60 years.

This research paper is organized as follows: Section 1 introduces the background and the context in which the study has been undertaken. Section 2 presents the relevant literature highlighting various ways by which research literature has been analyzed. Section 3 explains the methodology used in the study. Section 4 discusses critical research findings along with comparison of two data analytics platforms while some concluding remarks including major results are presented in Section 5.

## 2. Literature Review

This section presents a review of the literature on the application of text mining techniques in various domains whether it may be management, logistics or data analytics using computational Programming. The initial part deals with application of text mining techniques to the present aspects of data analysis for research literature while the later part presents various applications of text mining approaches in different domains.

Text mining has been utilized by various studies to analyze the methodologies and contexts of research trends of the innumerable articles published under IJPR journal solely concerning data analysis on a macroscopic level. The most recent study incubating text mining and big data analysis in marketing puts forward its idea of analysis based on relevant terms and topics related with five dimensions: Big Data, Marketing, Geographic location of authors' affiliation (countries and continents), Products, and Sectors, etc. (Amado et al., 2017). Some of the well-known research works in the contemporized situations have given new and meaningful definitions to research domains in this regard. Bag et al. (2017) tried to analyze e-commerce retailers and prepare an efficient search platform for the customers to obtain the desired durable goods in an efficient and effective manner. Sensitivity analysis has been commemorated in order in imbibe the potency of the proposed study model. Moreover, this paper has tried to integrate the in skirts of market studying and its effects on the customers across the globe, initiating its potency.

Research in big data applications is still in an embryonic stage. Hence it is quite clear that more firm efforts need to be put on in order to bring about majestic transformations for Big Data to thrive in the Marketing arena. However, in regard to it, still some great works have been forfeited and still a lot more needs to come. Jian Jin et al. (2016) comprehended big data analytics focusing on majorly 4 V's (Volume, Variety, Velocity and Value or 4 V's and this indeed projected the specificity of the paper in a more subtle and analytical manner. The main be all and end all here is to comprehend the demands of the users via mainly focusing onto help designers to understand the changes of and their competitive advantages. Large packet data is then fragmented out in whichever manner required initiating the utility of big data analytics.

The roots of text mining stretches back to late 1995's where the fundamental difference between 'text mining' and 'data mining' was still being apprehended. Many past researchers have had tried to analyse the applicable difference between the two but little have either being successful before 2004's. However, Written (2004) demarcated many major differences between the two. The phrase 'Text Mining' is used to generally denote any type of system that analyses large quantizes of simple sober language in the form of text and detecting the lexical terms to decode useful essential information efficient for research growth (Sebastiani, 2002). In discussing a topic that lacks a generally accepted definition in a practical Handbook such as this, Written chose to cast the net widely and take a liberal viewpoint of what should be included, rather than attempting a clear-cut characterization that will inevitably restrict the scope of what is covered, In this manner text mining is a proliferating technology incorporating various entities such as document classification, entity extraction and filling templates that corresponds to given relationships between entities, so that these can most mostly aggregated in the future perspectives.

The history of text mining right from the advents of 2000 to late 2010's has developed infinitely many domains for analysis of research data and its basic applications in our daily lifestyle (Elder et al., 2009 The major goal of text mining is based on various perspectives such as Identification of sets and clusters from the available reports and then finally re-structuring the textual information to discover hidden patterns that could provide some useful insights reacted to causes of fatal accidents and identification of frequent item sets etc.. This has indeed provided a series of benefits for ongoing study and its future prospects in the coming years. It can be applied to many applications in a variety of fields, namely, marketing, national security, medical and biomedical, and public relation. This knowledge of text mining has many optimistic approaches residing under it such as further studies done in research mining are actually an aid to the past research analysis.

Thus, the application of a text mining approach for analyzing the data in textual form presents huge advantages, which foster innovations that have economic and social benefits. Katsaliaki et al. (2014) meticulously used metadata and citation analysis to profile OR research and practice published in this prestigious journal. This paper aims at profiling OR research and practice published in JORS in a span of 10 years (2000-2009). Reviewing and profiling existing JORS publications helped to identify currently under-explored research issues, and select theories and methods appropriate to their investigation, all of which are recognized as important issues for conducting fruitful, original and rigorous research. Similarly, Kavakoitis et al. (2017) tried to present the aids relating to text mining for ailing against Diabetes embracing its significances. The main objective of text mining approach was vividly based the key goals of reviewing of applications of machine learning, data mining and workings in the field of medicines especially ill wills such as diabetes under Health Care And Environment. Here the topmost priority will be 'Prediction and Diagnosis'. A wide range of machine learning algorithms were employed which indeed helped in generating many possible evidences for combating Diabetic Disorders.

Kapoor et al. (2017) tried to understand the advances in Social Media Research and information systems identifying multiple emergent themes in the existing corpus within their field of research. A prominent output of this research paper is the generation of massive amounts of information, offering users exceptional service value proposition and robust intellectual framework for social media research that would be of immense value to academics and practitioners alike. Moreover, Priyani

(2017) tried to introduce Computational Mapping of OMSA (Opinion Mining and Sentiment Analysis, a relatively new but fast growing research discipline integrating the use of textual Mining along with research analysis. Indeed the paper presents a detailed analytical mapping of OMSA research work and charts the progress of discipline on various useful parameters.

Pablos et al. (2017) and his team of researchers ssupervised approaches for Aspect Based Sentiment Analysis. This study indeed helped them to achieve good results for the domain and language they are trained on, however lot of setbacks too incorporated in between as the quantity of research analysis could not be realized on larger segments of data available. This was majorly due to manually labeling data to train supervised systems for all domains and languages and this indeed proved to be very costly and time consuming as well. However, the result out casted from the study helped to imbibe important aspects, which ultimately attracted customer and utility services across the globe.

Apart from this, many other disciplines have also been studied forward such as IOTs, Web Mining, etc. and lot of works has been accomplished in this regard. Table 1 presents various other studies done in past one decade in the field of text mining:

Table 1: Literature Summary for Text Mining Applications in Various Domains

| S.No. | Authors | Domain | Findings |
|---|---|---|---|
| 1 | Kusiak (2011) | Data Mining: Manufacturing and Service Applications | ● Well Defined framework for organizing and applying knowledge for decision-making in manufacturing and service applications is presented.<br>● Examples of data mining applications in industrial, medical, and pharmaceutical domains are presented.<br>● Future perspectives of the paper include the envisioning of the data provided to exemplify future goals. |
| 2 | Psannis et al. (2014) | Convergence of Internet of things and mobile cloud computing | ● IoT-enables business services to highlight the benefits of converging IoT and MCC.<br>● Immense Benefits of Cloud-Computing incorporating IOT technologies.<br>● Convergence of green IoT and MCC framework well define the future goals along with focusing on Limitations of IOT in the present scenario. |
| 3 | Shu (2016) | Big Data Analytics: Six Techniques | ● Collective gathering and Individual generation are two common sources to procure data.<br>● Six Techniques being; Ensemble Analysis, Association Analysis, High Dimensional Analysis, Deep Analysis, Precision Analysis and lastly Divide and Conquer Analysis.<br>● Limiting the scope of data analysis due to lack of proper advancements in technology. |
| 4 | Ahmed et al. (2017) | Role of Big Data Analytics in Internet Of | ● Focus of IOT Applications on monitoring discrete events and mining the information collected by IoT objects. |

| | | Things | ●Existing big data solutions in the IoT paradigm still in their infancy wherein the associated challenges needs to be addressed. |
|---|---|---|---|
| 5 | Esteban (2017) | Onto Gene Web Services For Biomedical Text Mining Applications | ● Navigating large quantities of text within a range of different biomedical settings.<br>● Biomedical Applications processed in the following areas hospitals, academic laboratories, government safety and regulatory agencies, and pharmaceutical research and development.<br>●Figure Mining and Patent Mining Applications pivotal to text mining and its various perspectives. |
| 6 | Wu et al. (2017) | Implementation of Web Mining Algorithm Based On Cloud Computing | ● Study constructs a distributed cloud environment using a Hadoop framework.<br>● Setting the different block size parameters to amplify the level of performance.<br>● Block size determines the number that the pending data file is split, and the corresponding scale and amount of parallel calculation. |

Thus the applications of text mining approach for analyzing and interpreting large stocks of unstructured data explicit many advantages (Amado et al., 2017) some of those are as follows:

- Efficiency in Performance: the text mining approach allows for the efficient analysis of extant literature, this helps to validate large stocks of raw information available in least period of time.
- Interpretation of Useful Information: the analysis also helps with the extraction of useful information hidden in large numbers of research publications and presents underlying connections between various topics.
- Unrevealing Innovative Research Questions It helps to unfold much extremely important information effectively using enhanced tools and frameworks necessary to improve learning processes in this complex and interconnected world.
- Re-Defining research processes: With the growth of text analysis incorporating new Modified research processes that integrate text-mining tools not only explore new horizons, they also help to map results within specific contexts.
- Other benefits: in addition to benefits, as mentioned above, this approach also helps to maximize productivity gains with huge cost savings and resultant new business models. Better selectivity of the accomplished approach to provide validated responses in the coming future are some of its major assets being implemented.

This section has presented a review of relevant literature relating to the aspects of text mining approaches in big data analytics. It can be inferred from the above review that with growing data, it is essential, though quite difficult, for organizations to analyze extensive information that is kept in textual formats. To be able to extract hidden information from text, data mining techniques are most suitable. Though text mining is a new phenomenon, however the applications of text mining incubating various fields are innumerable. In this study, an attempt has been made to analyze the suggestions obtained from a survey of both researchers and industry professionals that was intended to enhance aspects of text mining along with the extensively comparable use of Python and R programming. In addition to this, the study also presents the similarities and differences in the opinions of both the groups with respect to what they understand of each other's concerns for collaboration. Hence, the main objectives of text mining approach inculcate the advancement and growth in its applications using the realm of researchers and professionals for extracting the hidden encrypted information for future endeavors.

Thus, this research paper is intended to answer the following four important research questions using text-mining approaches within the context of exploring big raw unstructured data using computational programming.

1. RQ1. Encoding research citation data for realizing the goals of text mining approaches and its varied applications?
2. RQ2. What are the methods to study and analyze the processed data after structuring based on Python and R Programming platforms?
3. RQ3. Which important words are emerging in different clusters and what are the topics for each cluster based on both platforms?
4. RQ4. What Conclusions can be drawn from the topics framed based on literature and which programming platform is more effective in analyzing the research data?

## 3. Methodology

In this study, an attempt has been made to explicitly describe the various steps executed to collaborate the research data by analyzing various text mining approaches in a sequential manner, such as data collection, data preparation, and data clustering and finally followed by a picturing of the whole scenario using data visualization. The study has specifically focused on comparison of Python and R Programming language to analyze textual data using a web based tool namely NLTK i.e. Natural Language Toolkit for converting the raw available data into useful tokens and then finally visualizing the word frequencies using Word Clouds. There basic difference between R and Python lies in the fact whereas R focuses on better user-friendly data analytics, statistics and graphical models Python on the other hand relies heavily on emphasizing productivity and code readability. However based on our studies both the languages presumably tend to follow same approach of passing on to a number of steps as presented in Figure 1 below:

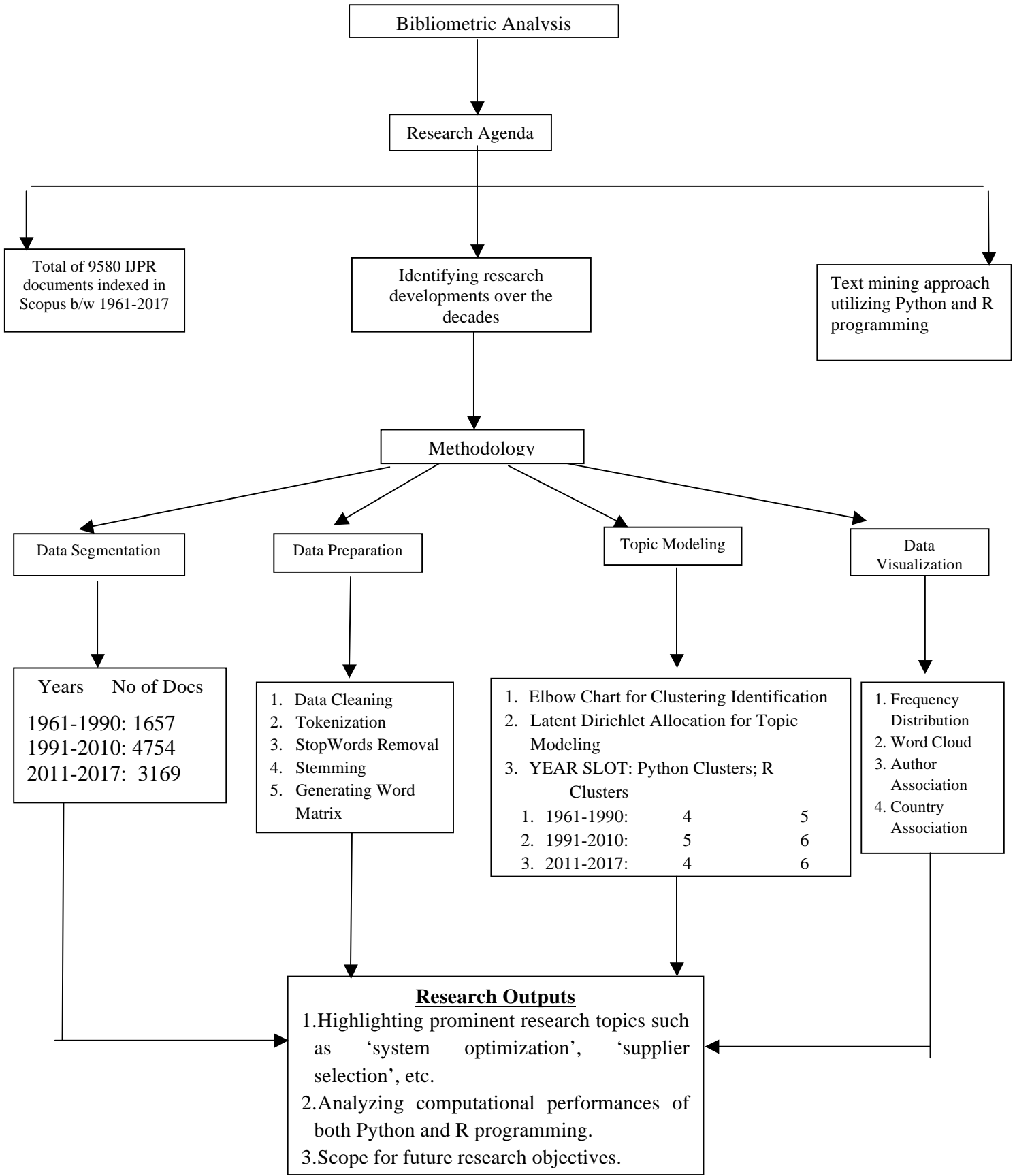

Figure 1. Methodology for Identifying Research Trends

### 3.1 Data Collection

The be all and end all of our research study has dwindled across IJPR Journal to highlight various research trends in different domains available for further studies. Using Python and R programming, the research study became more adaptable owing to the use of 'bib' format which includes important citation records such as name of authors, paper title, abstract, publication year, affiliation, etc. While extending our work force, limitations of IJPR also came across us. Such as the problem of not fetching full entity of information from the various sources available as at a time, citation record of only 1 paper could be downloaded. Now for carrying out such a huge research on thousands of research papers entitled in the journal, it would have been extreme cumbersome and time-consuming. Hence we conceptualized our idea by switching over Scopus for further research analysis. In Scopus, initially we found 9580 articles of IJPR journal as on 12 November 2017. Further, it came to our notice that the citation record of at most 2000 papers can be downloaded at a particular time. Hence, to fit to the needs of Scopus, we again calibrated our approach by dividing our research database into 3 major segments over a span of 5 decades namely from 1961-1990, 1991-2010 and finally from 2011-2017. The major benefit provided by this modification helped us to realize the main aim of implementing our text mining approach over a range of selected research papers published in IJPR and indexed in Scopus, ultimately providing mind-boggling research trends from our research studies.

### 3.2 Cleaning Of Data

The foremost step in research study incubated collecting data from various feasible resources. For this, IJPR was strictly targeted with the viewpoint to study the various available works accomplished till then and developing various text-mining approaches for fully grasping the essence of big data analytics using python and R programming. After collection of data its cleaning in a well oriented manner is the topmost priority.
 The various steps confronted during cleaning of data are explained as below-:

3.2.1 Transforming raw data: in this step, the ambiguities present in the research data collected from vibrant sources are first tackled and then finally put forward in a structured form to convert them into a suitable format required by the Python and R tool in order to consider each suggestion as a different document and analyses it accordingly.
Thus, in this manner, the responses obtained from researchers and industry professionals were transformed into the required format separately so that comparison of the results could be made suitably.

3.2.2 Tokenizing data: It is the process of dividing string into small sub-parts in the form of relevant words, strings, keywords, etc. which are referred to as "Tokens" in computational programming. These are majorly separated either by whitespace or punctuation marks depending on the needs and specifically with regards to subject matter. Having impregnated with word vectors these tokens are deployed to filter usable words by applying relevant constraints.

3.2.3 Removing stop words and unwanted words: For highlighting the pivotalness of important words in text mining algorithms, removing of stop words becomes crucial for further studies. Stop Words such as Are", "were", "they", "that", etc., along with some additional unwanted words is also provided in the coding file. These according to the fields of usability are compared together and then finally unwanted irrational words are removed.

3.2.4 Stemming: Stemming is the process of reduction of word or even a phrase to its basic stemmed or 'rooted' form. The most potent use of stemming incorporates the significance of it in helping to redefine a small phrase or word to its fullest form depending wherever it may be implemented. Stemming most importantly removes suffixes from root words such as "ed", "ing"," ly", etc.

3.2.5 Generating word × document matrix (DTM): A Document Term Matrix most importantly helps to describe the frequency of terms that occur in a collection of documents. There are various schemes of finding out the possible values to be put out in the field of natural language processing. This is the final step of data preparation use matrix as the input for applying the text mining approaches and useful information decoded as an output for further analysis and study.

**3.3 Data Visualization**
The approach for extracting useful information from textual data using visual aids is referred as Data Visualization. This technique helps in extracting any complex events since they incorporate visual, phonological or even animated inputs in addition to textual formats to present data in multimodal ways (Kitchens, 2014). Following approaches have been taken into consideration for visualizing the IJPR Literature:

3.3.1 Author Association
The study of research data incubates large compiling of research material and extensive analysis relating to the project considered. This encompasses a list of authors having made a globule of contribution in different fields of research analytics. This indeed many a times exemplifies that the same author becomes a common benefactor in many similar research accomplishments (Shastry et al., 2016).
In other words, the author associations are generated as a result of multiple collaborative works produced by different author relating to the same field of work in IJPR. This research paper presents transparent works of author associates contributing to the research area under study. The validation of the text, visual and factual data inherited from different sources has been sincerely studied and then put forward in a more compact yet objectionable manner.

3.3.2 Country Association
The best part of our study was that it was successful in bringing out the association of various country researchers working in coordination under a given time frame and bringing out new innovative research works in collaboration with different country researchers.

For this various approaches were confronted using Python programming helping us to extensively find out the contribution by different country researchers to the ongoing paper works and research studies.

### 3.3.3 Frequency Representation

Graphical Representation always aids textual and factual projection of data as the former edges against the latter by providing better comprehension and research- mindset to the readers (Gafarov et al., 2014). The consequences of all such processes are that the modern research papers find extravagant use of frequency representation and its varied tools in their research interests.

Our research area has enormously represented the data incurred after stemming processes during the cleaning period in terms of their repeating behavior and special attention has been focused on trends incubated after fetching top stemmed words along with the years in which study has been focused into.

### 3.3.4 Year Frequency Graph

This paper indeed tries to intensify the research study via incorporation of text mining steps for particularly dividing the research work into 3 main fragments of years starting right from 1961 till 2017. After the cleaning process has been successfully accomplished, words obtained after stemming corresponding to any of these 3 categories of years are represented on a graph plotted between frequencies of the stemmed words along with their years. In this way research analysis becomes more approachable and structured when collaborated together along with the other journals relied to make this paper realize its ultimate objectives.

## 3.4. Topic Modeling

In this contemporized world where research data is available in unstructured format it's very important that data is analyzed in a selective format suitable for growth and development (Xu Yin et al., 2017). Adding to the same quality, technology has developed some powerful methods, which can be used to mine through the data and fetch the information that we are looking for. One such technique in the field of text mining is Topic Modeling (Tomberg et al., 2016). As the name suggests, it is a process to automatically identify topics present in a text object and to derive hidden patterns exhibited by a text corpus. Thus, assisting better decision-making (Brasethvik, 1998).

Topic Modeling is quite different from rule-based text mining approaches that use generally outdated techniques of searching in formations or expressions presented accordingly. It is an unsupervised approach used for finding and observing the bunch of words (called "topics") in large clusters of texts. Topic Modeling is useful in innumerable cases such as documentations and clustering, organizing large blocks of textual data, information retrieval, etc. can easily be studied (Tatersall et al., 2006).

### 3.4.1 Elbow Method

After word - acquisition, title handling comes into picture. For this study, both R and Python have built the same text mining approaches for calculating the variation of words using Elbow Charts (Viswanathan, 2017**).**The variations obtained using functions as clusters are mainly analyzed by using the basic Elbow Method. One should choose a number of clusters so that adding another cluster doesn't give much better modeling of the data. If one plots the percentage of variance explained by the clusters against the number of clusters, the first clusters will add much information (explain a lot of variance), but at some point the marginal gain will drop, giving an angle in the graph. The number of clusters is chosen at this point, hence the "elbow criterion". This "elbow" cannot always be unambiguously identified.

3.4.2 Latent Dirichlet Allocation (LDA)

In the most generalized approach of work where research observations needs to be explained by unobserved groups enumerating different research works that why some parts of data are similar and others different. In the most simplistic forms, Latent Dirichlet allocation (LDA) is a generative statistical model that allows ambiguities present in the intended data to become less as due to grouping a particular pattern becomes easy to identify and even abide to (Kakusazde, 2017).

LDA technique is the most popular topic modeling technique due to easy implementation and it is relatively fast too. The distance between the observations is measured and is finally grouped together into a suitable number of clusters based on the closeness of the data. In this study, Gibbs sampling method has been used for fitting topic models. When Gibbs sampling is used for fitting the model, seed words with their additional weights for the prior parameters can be specified in order to be able to fit seeded topic models (Masood et al., 2017)

The steps for applying of LDA algorithm are as follows (Blei et al., 2003):

1. LDA makes assumptions that documents are produced from a varied extent of topics. Those topics generate words based on their frequency distribution and variations. LDA mostly incorporated backtracking facility figuring those topics created from documents in the first place.
2. LDA is a matrix factorization technique. This makes it most extensively feasible and easy to be used for in the vector space. The document term matrix created from corpus gets converted into two lower dimensional matrices – M1 and M2. However, these distribution needs to be improved, which is the main aim of LDA. LDA makes use of sampling techniques in order to improve these matrices.
3. All the text documents combined is known as the corpus. To run any mathematical model on text corpus, it is a good practice to convert it into a matrix representation. LDA model looks for repeating term patterns in the entire DT matrix.
4. Next step is to create an object for LDA model and train it on Document-Term matrix. Training incubates various explained processes already discussed above. The gensim module allows LDA estimation in an effective manner and this ultimately makes LDA implementation a very simple and much popular among researchers.
5. Characteristic Features -- Both K-means and Latent Dirichlet Allocation (LDA) are **unsupervised learning** algorithms, where the user needs to decide a priori the parameter *K*, respectively the number of clusters and the number of topics.

If both are applied to assign *K* topics to a set of *N* documents, the most evident difference is that K-means is going to partition the *N* documents in *K* **disjoint** clusters (*i.e.* topics in this case). On the other hand, LDA assigns a document to a **mixture** of topics. Therefore each document is characterized by **one or more** topics (e.g. Document *D* belongs for60% to Topic *A*, 30% to topic *B* and 10% to topic *E)*. Hence, LDA can give more realistic results than k-means for topic assignment (Hoffman et al., 2010).

3.4.3 Word Cloud

After the accomplishment of network analysis using elbow method we now look forward on extensively using Tag Clouds. A tag cloud (word cloud or weighted list in visual design) is one of the most efficient methods of representing data in visual format amending it from textual data, these are predominantly used to depict tags on websites, which are usually single words, and the importance of each tag is shown with font size or color (Provan et al., 2011). The ultimate realm here is for quickly perceiving most dominant terms and for locating a term alphabetically to determine its relative prominence. The terms are hyperlinked to items associated with tags and thus the whole process gets executed.

From our study perspective, we have tried to realize the sole objectives in our research study by providing well defined pattern for analysis of word clouds citing the relevant information in a span of 5 decades and hence making the visual cumbersome and extensive data more appealing for understanding in a more subtle manner. Not only the research trends have been targeted but also its scope has opened infinite doors for coming researches based on the same benchmarks (Lohmann et al., 2014).

**4. Results and Discussions**

This section presents a schematic approach for applying text mining techniques to real-time situations incorporating the extensive use of Python and R programming and also comprehending the bulky outrageous data into sequential order using both languages and finally drawing relevant conclusions as to which of them suits the best for further research analysis. Finally the conclusions reimbursed from the research practices will help to define a theoretical modeling applicable to many research applications in the forthcoming years.

4.1 Overall Publication Trends

The total count of papers studied mainly under 3 major stretches was taken into consideration for extracting and focusing our attention towards a specified period of study. It includes trends incubated from 1960-1990, 1991-2010 and 2011-2017. The reasons for dividing the year stretches in 3 major slots have already been studied in great depth in the prior sections of the paper. As visible from research dataset; the minimum number of documents under study is somewhere around 1967. This trend is quite unclear due to varying highs and peaks in the research output analyzed during the whole stretch. 1Similarly, the maximum number of documents under study came around 1990 where the number of documents count was raised up to 158 (approx.) in a gradual manner since 1960 (Figure 1.1, Appendix A). The overall trend presented during 1991-2010 was very much similar to that presented between 1961-1990 (Figure 1.2, Appendix A). However the growth in trend is quite varied as may be clear from the figure that the stability set up in this case is quite less to what it was before and there are sudden uneven bumps too in the following data, owning to the growth and development of a new era in which the broad aspects of research was more glorified and had impregnated its way into the day to day research happenings and innovations.

The final slot of the years from 2011-2017 were analyzed using Python Programming and the trend showed quite uneven results with difference in the document count increasing exponentially right from 2012 to the following consecutive years (Figure 1.3, Appendix A). To conclude, the generalized type of result is almost same in all 3 but small changes can be neglected.

4.2 Frequency Distribution

One of the most potent results incorporated from our findings includes the trends related to frequencies of different words out casted from our study. While some potent words have occurred almost a dozen of times whereas the frequency of some words have been unevenly distributed throughout the research papers. This has ultimately been helpful in explaining the potency of the words and impregnating their importance throughout our study.

Python platform has been more optimizing and user-friendly in context to presenting the results from our research practices. It can be easily visualized using the python platform that the words such as "Group" and "Production" have occurred more than 1000 times during 1961-1990 while, "Process", "System" and "Manufacture" has occurred more than 3000 times during 1991-2010. The peaks in the graph quite evidently explain the situation of this research paper (Figure 2.1, Appendix B).

The frequency bar charts presented by R programming synonymously explains the frequency of various words over a given time slots (Figure 2.2, Appendix B). Some of the most predominant words existing in the literature over the different year slots have been found to be "problem", "procedure", "propose", "optimization", "customer", etc. For example, one major focus of most research papers have been "to sustain production methods as part of their problem statements and incorporate procedures for developing more stringent efficiencies for mass production by cutting cost and operation investments".

Comparatively, both Python and R programming have successfully provided us well – defined bar charts. However, R Programming has paved its way through due to well – detailed descriptive and comprehending way of data- presentation and allocation.

4.3 Word Clouds

The best way of representing the visual data is using Word Clouds extensively used during our research analysis. Words Clouds for the 3 different periods of time have been presented individually, integrating a major Overall Word Cloud for both Python and R programming giving us a more simplified and clearer approach over the period of time.

Python programming has successfully expressed word clouds more transparently and bringing about major dominating words during the following period of analysis. The word cloud have been presented in 3 different year slots, from 1961-2017 and in each, representation has been solely on the basis the size of words evolved during our research analysis. It also explains the significance of enlarged words such as group, system, supply chain, etc. via their specified amount to which they have been repeated in our analyses of data (Figure 3.1, Appendix C). This has helped us incite realistic practical details regarding the research pattern subsided during the following years of studying.

Words presented by R programming glorify important words synonymous to python programming incorporated during the same period of study. Here, the major focus of using R programming is to justify the performance stake of both the platforms on dealing with them differently (Figure 3.2, Appendix C).

From both the following results obtained here we further see that, the overall Word Cloud presented by Python programming is much more suitable due to its expressiveness and clearance in exemplifying the deep and vast information visually using the above method. On the other hand Python is a better approach, as it not only presents data meaningfully but also the research analysis generated.

Obviously, this might help researchers in the near future who would rather commemorate to presenting research data visually using Python programming rather than R, as already discussed in the above paragraph.

4.4 Author Association

There is always an author association generated to connectify the ongoing and already published works within different fields of study for exactly fetching proper information regarding each researcher and his works relating to a specific field of study.

Our study of research data provides an overall research association in a major 3sets of studying as remarked earlier. Here, the main purpose is to present the linkage and networks a particular author presents in his/her research and work analysis. The initial network includes all the authors while in the case of second of network; only those authors have been considered whose research contribution is more than 1%.

The author association developed with different entities over the time period is explained very subtly by Python programming over different years. The consequences of the results are different as can be inculcated from the graph. The part (1) explains the connections and networks generated by each author depending on the amount of common networks produced by him as evidently explained by dense lines in comparison to those of the part (2). In part (1) of the figure, the token number (109) (Author Name is Ben Arieh) represents most centrality in comparison to other authors as may be clear from above. This infers to the very fact that the research works presented by this author is most prominent in comparison to other authors. Similarly in case of second figure for authors as mentioned clearly by their token number such as (234), (373) enjoys maximum centrality and clearly most research inputs may be seen from above, names of these authors are mentioned as below – Dudley, Bullinger, etc. (Figure 4.1, Appendix D).

The author networks consisting of linked units have been explained with each network amount to explain the amount of connections generated by the author. The major difference between the two different slots is definitely the amount of research work done from 1991-2010 is much vast and deeper in extent contributing to greater number of network diagrams and therefore resulting in much greater number of connections as compared to the slot from 1961-1991.As clear from figure (1) Authors such as Khumawala (87), Mahmood (79), etc. emerged as maximum contributors during the period from 1961-1990. Similar results were brought about from Python Programming, namely the author Liang (147) established maximum centrality in comparison to others researchers during the aforementioned years (Figure 4.2, Appendix D).

Finally, the slot from 2010-2017 have been developed exemplifying excessive cluster of networks made by authors especially the centralized ones showing the extent and vastness of the study and research papers been published in the journal in the recent years (Figure 4.3, Appendix D).

Here, a major disadvantage incorporating the use of R Programming have been developed as R Programming could not display Author Associations as have been analyzed for Python Programming.

4.5 Country Association

The research work carried out by different researchers during the above period has ultimately rooted to form significant connections due to extensive research analysis carried out during the aforementioned years. Consequently, there has been a multitude of advantages due to more commonalities developed between researchers and their subsequent ideas regarding the same.

The research work impregnated by different country researchers was successful in bringing about very consolidating results through various generated graphs from by Python and R programming, as mentioned in this section.

1961-1990 was a time period when interconnection between different countries was actually on the state of development and undoubtedly a period of absolute revolution in many prominent countries of the world. Evidently, it is quite clear and valid able about the following points from the graph (Figure 5.1, Appendix E).

- The magnitude of research work carried out in the above-mentioned years definitely less as compared to the slots to be discussed in the following years.

- The magnitude of connections is directly proportional to the size of the enlarged circles representing the state of interconnectedness between different nations.
- As clearly visible from the graph that India is seen to be at the extreme right side of the graph. Since 1961-1990 was a period of renovation in India's History considering the effects of post-independence and The Great Bengal Famine on the Indian Subcontinent, the research work incorporated in such a state was undoubtedly seemingly astonishing. India made joint research connections with some of the mightiest powers at that time such as United Kingdom, New Zealand, Netherlands, and Thailand, etc.

Now, as the years passed by, the amount of research work showed a steady increase as more and more countries gained independence as well as financial stability.

The following points can be inferred from the graph (Figure 5.2, Appendix E).

- The magnitude of research work carried out in the above-mentioned years is definitely greater in comparison to the slot from 1960-1990.
- The Country participation has certainly been increased in the given mentioned years as already discussed earlier.
- United States presented centrally of the slot show maximum connections with various nations of the world exemplifying its mightiness in the research work over the following years.
- Indian research also saw an exponential rise in the magnitude and quality of research being generated and collaborated over various countries of the world. India certainly held a position among the top 20 countries of the world publishing research in connection with more than 28 top nations of the world.

Lastly, for the final year slot developed from 2011-2017, the following points can be inferred as clear (Figure 5.3, Appendix E).

- Maximum amount of research work have been carried out between different country researchers and officials.
- The country participation has seen a stupendous increase as compared to other slots adhering to the very fact that the number has almost two-folded as compared to past years.
- United States, China and United Kingdom can be considered to be the 'King makers' in scrutinizing the research trends and papers in the past decade or so.
- Indian research work form has not seen much catastrophic advancements in their research trends from early 2010's. This has implied the effect of various foreign policies and relations with other continents affecting the amount of research work presented in this case.

4.6 The Elbow Chart

The Elbow method is a method of interpretation and validation of consistency within cluster analysis designed to help finding the appropriate number of clusters in a dataset. In addition to this, elbow method helps to visualize an exact picture of the data incurred in an algorithmic manner so that the visuality of the sets is well defined. Not only is this, the readability and presentability of the represented data is also enhanced using elbow method extensively.

As quite evident from the previous sections, the foremost step of analyzing the cumbersome data is using Python programming. The main aim developed using Python programming here is to study the number of clusters over a period of time v/s their inertia relationships as obtained by Python programming. Submissively, in each of the 3 slots a tentative relationship is being established between the number of clusters and the inertia (the probability of cluster analysis in a more generic sense) over a period of time interval using the LDA approach of studying and clustering data. Based on the above plots, it can be inferred that number of clusters for the year slots 1961-1990, 1991-2010 and 2011-2017 are 4, 5 and 4 respectively based on the python program (Figure 6.1, Appendix F).

On the other hand, the analysis by R programming is satiated on different grounds and research frames. As clearly visible, the objective and the status of incurring relationship between Python and R programming is almost same but surprisingly the results out casted by R programming is way too different. Majorly because the process and the data representation provided by Python is much object able and appealing as compared to R programming.

It can be again inferred that number of clusters for the year slots 1961-1990, 1991-2010 and 2011-2017 are 5, 6 and 6 respectively based on R Programming (Figure 6.2, Appendix F).

Obviously, it can be picturesque that Python is conceivably much more suitable in comparison to R programming based on two prominent grounds:-

1. The irregularities in the data type presented in case of R programming is much more in comparison to Python programming .As a result of which the data presented is not suitable to be used and analyzed for further research analysis.
2. Though R programming has been capable for forming large number of clusters in comparison to Python programming. However, the inertia effect is much more dominant and edgy in Python programming, which makes it finally the most, suited to be used for research analysis and data interpretation.

4.7 Clusters

The clustering is one of the important data mining issues especially for big data analysis, where large volume data should be grouped. The k-means method remains the vibrant source of clustering and its various modifications, because it still remains one of the popular methods and is implemented in innovative technologies for big data analysis.

4.7.1 Topic Labeling

Topic Labeling has aimed towards providing a well-defined amount of words obtained in each cluster signifying their probability and amount of usage over the period of years. Table 2 presents trends observed under 3 time slots based on Python programming as shown below:

Table 2. Prominent Words by Python Programming

| Topic No | 1961-1990 | 1991-2010 | 2011-2017 |
|---|---|---|---|
| Cluster 1 | Cost, system, control, station, time | Chain, supplier, firm, operations, strategic | Algorithm, time, machine, solution |
| Cluster 2 | Part, problem, machine, system, process | Product, system, dynamic, control | Selection, group, chain, operations, process |
| Cluster 3 | Group, time, cost, method, different | Operations, systems, group, machine, solution | Method, system, part, product, operations |
| Cluster 4 | Method, tool, force, tool | Product, system, time, machine, group | Cost, chain, decision, results, market, product |
| Cluster 5 | -- | Tolerance, surface, error, method show, features | -- |

Words such as method, product, cost, tools, etc. became more common and saw tremendous amendments in research-oriented sectors. However, during 1991-2010, the impact of operations and supply chain strategies on integration and performances became more prominent in the aforementioned year slot, optimistic changes in the field-enhanced chances of more research in this area of specialization. Researchers now focused more on the processes governing the system and operations incorporating within it. Even by the end of 2010's, now pattern shifted more towards 'optimizing research solutions' than seen in the earlier years and paved a way more towards part of problem based solutions and conclusive statements. In the contemporary years from 2011-2017, now more focus and efforts by researchers are put on towards understanding the algorithms applied in decision-making process for sustainable developments and growth in varied research segments of the era. Words such as Solution, decisions, results etc. came more into pictures and gradually the emphasis now is more on conclusive outcomes in research activities along with the scope of sustaining information related to specific area of interest. However, definitely what need to be achieved in this segment is undoubtedly miles ahead and more stringent efforts needs to be reframed in this interest. Such as the cost of production needs to be minimized along with Decision making to achieve sustainable supply chains in the context of location of production, distribution facilities, etc.

Similarly, Table 3 presents the trends observed under 3 major time slots based on R programming. The approach using R based programming was seemingly much different when compared with Python. The period of 1961-1990 saw drastic historical changes in research-oriented activities, which have been predominantly due to the governance of Industries in this era. An implementation of measurement system analysis for assessment of machines, procedure relating to work growth-using R based approach was more prevalent in these years. During 1991-2010, the emphasis using R based approach was more on techniques and methods constituting the system, supplier chain policies and operations were talked about and implemented in the form of research papers, flexibility and transparency in research based segments became more prominent in the following years, Magnanimous changes came about during the following years from 2011-2017, mainly as the inclination was more towards IOT based approaches and amendments in the research fields. Words incurred using R programming such Performance, quality, solution, approach, etc. now came out strongly mainly due to advancements in technology over the following period of time. For e.g. A major research paper during the following years, on analyzing challenges to Internet of Things (IoT) adoption and diffusion: An Indian context focused the collaboration of IOT's in a plethora of Research segments such IOT-health, IOT-NDN environment, IOT-deployment on Soil Systems, etc. bringing about the spice of augmented multi varied outcomes on a different aspects of our day to lives.

Conclusively, one needs to realize that the research analysis presented during the contemporary years have majorly focused on improving and introducing intuitive methods relating to different domains of manufacturing, industrial operations, supply chain initiatives, etc. and a lot of that needs to achieved and enunciated in this regard.

Table 3. Prominent Words by R Programming

| Topic No | 1961-1990 | 1991-2010 | 2011-2017 |
|---|---|---|---|
| Cluster 1 | Production, method, time, procedure, machine | Chain, supplier, firm, operations, firm, cost | Method, present, strategy, relationship, integration |
| Cluster 2 | Work, system, order, tool, research, experimental | Product, system, control, dynamic, process | Manufacture, level, model, data, research, analysis |
| Cluster 3 | Result, solution, describe, optimal, computer | Performance, propose, result, control, system | Case, optimal, compare, effectiveness, operational |
| Cluster 4 | Group, approach, output, industrial, characteristics | Heuristic, technique, variable, set up, flexible | Problem, product, performance, group, chain |
| Cluster 5 | Jobs, material, component, simulation, analyst | Production, problem, management, improvement | Production, quality, solution, assembly, processes |
| Cluster 6 | -- | Machine, time, level, development | Algorithm, propose, supply, approach, result |

4.7.2 Clustering Trends

During our research analysis relating to clustering, there were various trends followed by clusters over a period of time, rejuvenating the very fact that Python indeed was better approach for providing well build cluster graphs over a range as discussed below.

4.7.2.1 Clustering Trends by Python Programming

All the clusters presented by Python programming have produced the same type of results. However, Cluster 1 sees an explicit growth over the years especially around 2010's and around. The amount of document has enormously increased during the following years. Except this Cluster all others have shown a steady decline in their document count over these years.

The amount of unevenness in the analyzed data after studying comes from during 1960-1991. (Figure 7.1, Appendix G). In these years the number of document count has strategically been dominated by many ups and downs in the following years. Also, it is quite very well evident that in the past a decade or so, the amount of document count has shown a steady increase especially from 2015's till the contemporized time.

In the year 2016, a maximum of 105 documents have been counted for research analysis and further studying. Optimistically, python has been successful in drawing pre-eminent citations and scope of research analysis that can be carried out in the following years to come. (Figure 7.2, Appendix G).

4.7.2.2 Clustering Trends by R Programming

Another, approach of incurring research analysis is via R programming. The results inculcated from R programming are majorly in the form of bar graphs and histogram types.

Cluster 2 exhibits maximum growth over the declining years and especially around 2010's. All other clusters have presented the data in similar manner as presented in the year slot from 1961-1990 specifying that there has been balanced approach in all the areas and clustering over these following years. (Figure 7.4, Appendix G).

From 1960 onwards, the trends exhibited by R programming could not establish well-defined peaks based on their documents counts. This indeed explained the constant increase of documentations and clustering over a range of years (Figure 7.5, Appendix G).

From 2011-2017, the amount of documentations has increased tremendously following an inconvincible exponential growth due to accelerated research studies as evident from the graphs shown above. Categorically scrutinizing, both python and R programming prove handy and accessible in developing Clustering Trends for 3 consecutive year slots. Whereas, R programming represents trends in the form of bar charts, Python programming does it by simple curve sketching and its canonical representations (Figure 7.6, Appendix G).

4.7.3 Cluster Wise Word Cloud

Cluster Wise Cloud is mainly developed to provide visual understanding of the graphical cumbersome data so that a plethora of research citations and conclusions can be framed out just by visually accessing data. The importance and usage of words varies according to their size in word cloud representations and thus helps users to analyze the global variations and type of work incorporated in a particular research paper.

4.7.3.1 Clustering Word Cloud Python

Clustering words clouds have been suitably presented by Python programming developed over a stretch of more 30 years or so (Figure 7.7, Appendix G). Following are the inferences that can be drawn from the various Clustering Word Clouds shown above:

1. From 1961-1990, words such as 'cost', ' system', ' station', ' method', etc. became more prominent and exhibited maximum frequency. Similarly, from 1991- 2010 words such as 'chain', 'supplier', 'operations', ' firm', etc. saw an increased rise as evident from the word cloud shown above.

2. Finally in the year slot from 2011-2017, technologically driven terms such as algorithm, solution, time, machine, etc. became more concrete and dominant to use. These have helped us to build well-structured definitions of the kind of work, its potency and its cognizance in relation to specific domain such as Sustainable Manufacturing, Supply Chain Management etc. in relation to IJPR journal.

4.7.3.2 Clustering Word Cloud R Programming

Similarly, Clustering Word Clouds have also been developed by R programming during the 3-year slots (Figure 7.8, Appendix G). Following inferences can be drawn from the Clustering Word Cloud in R programming:

1. Obviously the results shown by R programming in the 3 major year slots is in par with the results presented by Python Programming, adding to the very fact that the process of data representation is way too different to those of Python programming.

2. From 2011-2017 words such as 'method', 'constraint', 'strategy', etc. became increasingly important and benevolent. These helps to crisp out the importance that more focus of most research papers during the following decade was to optimize cost of produced items, their operations and most prominently devising new methods and strategies for developing broad perspectives for the same.

4.8 Time Taken in Different Computations

The time taken by different programming languages varied differently across different domains and under different slots of times periods as may be seen clearly. Different data can be analyzed for both of them as may be evidently visible in the sub sections discussed below-:

4.8.1 Data Analysis by Python Programming

The results by Python programming over 3 different slots varied quite differently. Moreover, the approach of data presentation by Python is way too different as to that of R programming.

Table 4. Computational Performance for Python Program

| S.NO. | ATTRIBUTES | TIME TAKEN (in secs) | | |
|---|---|---|---|---|
| | | 1961-1990 | 1991-2010 | 2011-2017 |
| 1. | No. of Documents | 1657 | 4754 | 3169 |
| 2. | Tokenization | 4.697 | 18.202 | 13.18099 |
| 3. | Cleaning of Data | 405.151 | 6059.582 | 3574.473 |
| 4. | Formation Of Overall Trend | 17.666 | 1099.822 | 14.316 |
| 5. | Word Cloud Formation | 71.698 | 152.412 | 60.8700 |
| 6. | Formation Of Frequency Distribution (top 200 words) | 77.432 | 11.788 | 38.187 |
| 7. | Formation of Author Association | 63.305 | 74.048 | 64.927 |
| 8 | Formation Of Elbow Chart | 146.257 | 605.489 | 523.511 |
| 9. | Topic Formation | 758.650 | 1070.816 | 1045.013 |
| 10. | Formation of Word Cloud Of All Cluster | 64.013 | 32.269 | 30.891 |
| 11 | Network Of Topics | 32.914 | 27.0510 | 28.102 |

4.8.2 Data Analysis by R Programming

The results provided by R programming varied to those of Python Programming since the steps to incorporate the procedure were different to those of Python Programming. Following table shown below represents the time taken by R programming to procure the results as may be seen clearly from above:

Table 5. Computational Performance for R Program

| S.NO. | ATTRIBUTES | TIME TAKEN (in secs) | | |
|---|---|---|---|---|
| | | 1961-1990 | 1991-2010 | 2011-2017 |
| 1. | Corpus Formation | 2.338 | 3.708 | 1.571 |
| 2. | Cleaning Corpus | 6.454 | 59.187 | 14.335 |
| 3. | Elbow Chart Formation | 3.141 | 46.025 | 18.284 |
| 4. | Cluster formation | 9.596 | 1.691 | 1.003 |
| 5. | All the word cloud formation | 11.580 | 26.028 | 21.520 |
| 6. | Overall trend formation | 0.75 | 2.464 | 0.785 |
| 7. | All topic trend formation | 8.589 | 5.552 | 8.740 |
| 8. | Topic assigning using LDA | 13.796 | 1.060 | 44.811 |

4.9 Comparative Analysis

As can be easily perceived from both the results, that a well–defined comparative analysis can be stretched from both the programming languages executions and incorporation-:

- The steps used for analyzing data in both Python and R programming is certainly differentiable but as can be seen clearly that the clarity of visually represented data turns out to be much more in case of Python language. Ultimately, Python language is much faster and executable in comparison to R programming as in various cases the time taken by Python programming is certainly much less in comparison to R programming.

- Python based approach differs from that of R programming much because of their difference in analyzing system Approach on the grounds of research oriented activities. In the contemporary period, Python has brought about technology driven along with substantial amount of information from the data available whereas R programming has relied much towards tech-oriented advancements and this has indeed created a major difference, in both of their ways of handling raw unstructured data.

- Certainly the GUI (Graphical User Interface) provided by python programming for running codes or analyzing data is much more prominent is terms of presenting data along with user-friendly orientations. Some of which are undoubtedly lacking in R programming which indirectly affects both platforms computational performances.

**5 Conclusions**

As rightly stated the potential of any research is measured by the results out casted by it based on the subjective and comprehensive study done on research data and its various facets. Ultimately the goal of analyzing big cumbersome data into substantial fragments having been fulfilled by Python and R programming as per the results discussed in the given sections leads us to draw synchronizing conclusions based on our research analysis. This indeed has provided validating answers to the various research questions mentioned under methodology in a more subtle manner. Research citations' after being encoded has turned majestically to provide stupendous results to realize the ultimate objectives of text mining as clearly stated here.
A series of well-stated differential outcomes can be brought about in the manner; Python and R programming have impregnated their approach of handling raw unstructured data over the 3 major year slots. Right from 1961 onwards when mechanization and industrialization were setting its feet forward, Python framed out a strong analytical approach of understanding the raw uneven data available at that time, on the other hand R programming focused more on implementing system analysis on assessment of machines and varied procedures for re-building research outcomes for the specified period of time. This is small citation of indeed how Python and R programming have differed in depth on the grounds of analysing system approaches for bringing about different conclusions over a period of time.

Ultimately the goal of understanding the text mining approach over a period of time using varied methodologies has been able to develop well oriented outcomes using both Python and R programming however, Python has developed an over-hand in handling situations and dealing with problems much due to its influential executing speed, visual-clarifications and strongly comprehending raw suits of data from a plethora of research data available. Justifiably, this will undoubtedly open vast fields of research interests and future scopes for the researchers initiating in the same arena, by providing them the necessary know-how (prototype) as well as re-embrace their footprints in the same domain by extending to other databases also in addition to IJPR, to eject stupendous outcomes similarly in a plethora of research-dominated fields such as medical, management, archeology, etc.

**References**

Ahmed, E., Yaqoob, I., Hashem, I.A.T., Khan, I., Ahmed, A.I.A., Imran, M. and Vasilakos, A.V., 2017.The role of big data analytics in Internet of Things. Computer Network available at http://www.imrankhan1984.com/uploads/9/7/6/8/9768503/bigdataanalyticsiniot.pdf, Accessed On: 15 October 2017.

Amado, A., Cortez, P., Rita, P. and Moro, S., 2017. Research trends on Big Data in Marketing: A text mining and topic modeling based literature analysis. European Research on Management and Business Economics, Available at: http://www.sciencedirect.com/science/article/pii/S2444883417300268, Accessed on: 10 October 2017.

Bag, S., Tiwari, M.K. and Chan, F.T., 2017. Predicting the consumer's purchase intention of durable goods: An attribute-level analysis. Journal of Business Research. Available at: https://www.sciencedirect.com/science/article/pii/S0148296317304770, Accessed on: 10 June 2018.

Blei, D.M., Ng, A.Y. and Jordan, M.I., 2003. Latent Dirichlet allocation. Journal machine learning research, 3, pp. 993-1022, Accessed On: 2 November 2017.

Brasethvik, T., 1998. A semantic modeling approach to metadata. Internet Research, 8(5), pp. 377-386. Accessed On: 25 October 2017.

Chaudhry, S.S. and Luo, W., 2005. Application of genetic algorithms in production and operations management: a review. International Journal of Production Research, 43(19), pp. 4083-4101, Accessed On: 2 November 2017.

Gafarov, E.R., Dolgui, A. and Werner, F., 2014.A new graphical approach for solving single-machine scheduling problems approximately. International Journal of Production Research, 52(13), pp. 3762-3777.

García-Pablos, A., Cuadros, M. and Rigau, G., 2018. W2VLDA: almost unsupervised system for aspect based sentiment analysis. Expert Systems with Applications, 91, pp.127-137.

Heimerl, F., Lohmann, S., Lange, S. and Ertl, T., 2014, January. Word cloud explorer: Text analytics based on word clouds. In System Sciences (HICSS), 2014 47th Hawaii International Conference on (pp. 1833-1842). IEEE.

Hoffman, M., Bach, F.R. and Blei, D.M., 2010. Online learning for latent dirichlet allocation. In advances in neural information processing systems, pp. 856-864.

Hong S., 2016. Big data analytics: six techniques, Geo-spatial Information Science, 19(2), 119-128, Available at: http://www.tandfonline.com/doi/pdf/10.1080/10095020.2016.1182307, Accessed on: 5 October 2017.

Jin, J., Liu, Y., Ji, P. and Liu, H., 2016. Understanding big consumer opinion data for market-driven product design. International Journal of Production Research, 54(10), pp. 3019-3041.

Kakushadze, Z. and Yu, W., 2017. K-means and Cluster Models for Cancer Signatures. Available at: https://arxiv.org/pdf/1703.00703.pdf, Accessed On: 28 October 2017.

Kapoor, K.K., Tamilmani, K., Rana, N.P., Patil, P., Dwivedi, Y.K. and Nerur, S., 2017. Advances in social media research: Past, Present and Future. Information Systems Frontiers, pp. 1-28.

Katsaliaki, K., Mustafee, N., Dwivedi, Y.K., Williams, T. and Wilson, J.M., 2010. A profile of OR research and practice published in the Journal of the Operational Research Society. Journal of the Operational Research Society, 61(1), pp. 82-94.

Kavakiotis, I., Tsave, O., Salifoglou, A., Maglaveras, N., Vlahavas, I. and Chouvarda, I., 2017. Machine learning and data mining methods in diabetes research. Computational and structural biotechnology journal, 15, pp. 104-116.

Khan, Z., Khan, Z., Vorley, T. and Vorley, T., 2017. Big data text analytics: an enabler of knowledge management. Journal of Knowledge Management, 21(1), pp. 18-34.

Kitchens, M.B., 2014. Word Clouds: An Informal Assessment of Student Learning. College Teaching, 62(3), pp. 113-114.

Kusiak, A., 2006. Data mining: manufacturing and service applications. International Journal of Production Research, 44(18-19), pp. 4175-4191.

Masood, Muhammad Ali, et al. "MFS-LDA: a multi-feature space tag recommendation model for cold start problem." Program 51.3 (2017): 218-234.

Napoleon, D. and Pavalakodi, S., 2011. A new method for dimensionality reduction using K-means clustering algorithm for high dimensional data set. International Journal of Computer Applications, 13(7), pp. 41-46. Accessed On: 21 October 17.

Nisbet, R., Elder, J. and Miner, G., 2009. Chapter 21-Prospects for the Future of Data Mining and Text Mining as Part of Our Everyday Lives. Handbook of statistical analysis and data mining applications, Academic Press, ISBN: 978-0-12-374765-5, pp. 755-778. Accessed On: 13 November 2017.

Parmar, D., Wu, T., Callarman, T., Fowler, J. and Wolfe, P., 2010.A clustering algorithm for supplier base management. International Journal of Production Research, 48(13), pp. 3803-3821. Accessed On: 21 November 17.

Psannis, K.E., Xinogalos, S. and Sifaleras, A., 2014. Convergence of Internet of things and mobile cloud computing. Systems Science & Control Engineering: An Open Access Journal, 2(1), pp. 476-483. Accessed On: 26 October 2017.

Reynolds, W.A. and Cheung, M.K., 1984. The use of industrial engineering in Hong Kong manufacturing industry. International Journal of Production Research, 22(6), pp. 983-999. Accessed On: 10 October 2017.

Rodriguez-Esteban, R., Text Mining Applications, Available at https://doi.org/10.1371/journal.pcbi.1000597, Accessed On: 28 October 2017.

Sastry, C.S., Jagaluru, D.S. and Mahesh, K., 2016. Author ranking in multi-author collaborative networks. Collnet Journal of Scientometrics and Information Management, 10(1), pp. 21-40.

Sidorova, A., Evangelopoulos, N., Valacich, J.S. and Ramakrishnan, T., 2008. Uncovering the intellectual core of the information systems discipline. Mis Quarterly, pp. 467-482.

Tattersall, C., Janssen, J., Van den Berg, B. and Koper, R., 2006. Modelling routes towards learning goals. Campus-Wide Information Systems, 23(5), pp. 312-324. Accessed On: 15 October 2017.

Vishwanathan, K., Alizadehkhaiyat, O., Kemp, G.J. and Frostick, S.P., 2017. Minimal clinically important difference of Liverpool Elbow Score in elbow arthroplasty. JSES. Accessed On: 15 October 2017.

Witten, I.H., 2004. Text Mining. Available at https://www.cs.waikato.ac.nz/~ihw/papers/04-IHW-Text Mining.pdf, Accessed On: 30 October 2017.

Wu, Y., Provan, T., Wei, F., Liu, S. and Ma, K.L., 2011. Semantic preserving word clouds by seam carving. In Computer Graphics Forum, 30(3), pp. 741-750. Accessed On: 12 October 2017.

Wu, W., Chen, Y. and Seng, D., 2017. Implementation of Web Mining Algorithm Based on Cloud Computing. Intelligent Automation & Soft Computing, pp. 1-6. Available at http://dx.doi.org/10.1080/10798587.2017.1316077, Accessed: on 20 October 2017.

Xu, Y., Yin, Y. and Yin, J., 2017. Tackling topic general words in topic modeling. Engineering Applications of Artificial Intelligence, 62, pp.124-133. Accessed On: 28 October 2017.

# APPENDIX – A

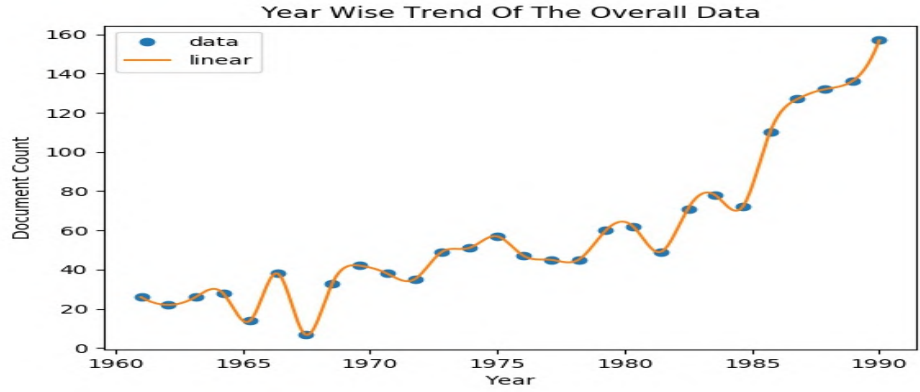

Figure 1.1 Year Wise trend of Overall Data for 1961-1990

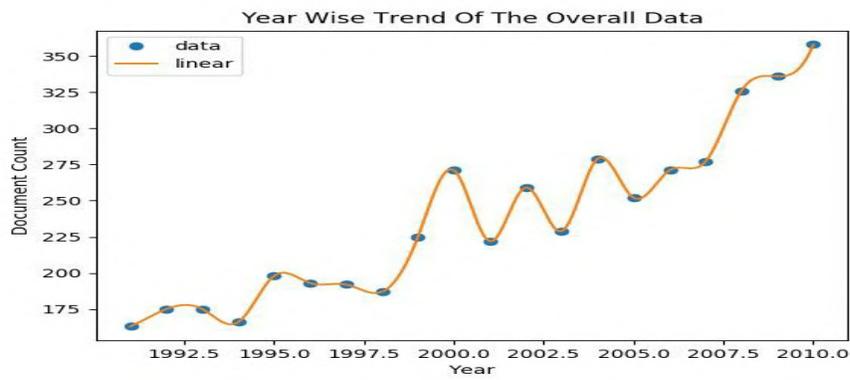

Figure 1.2 Year Wise trend of Overall Data from 1991-2010

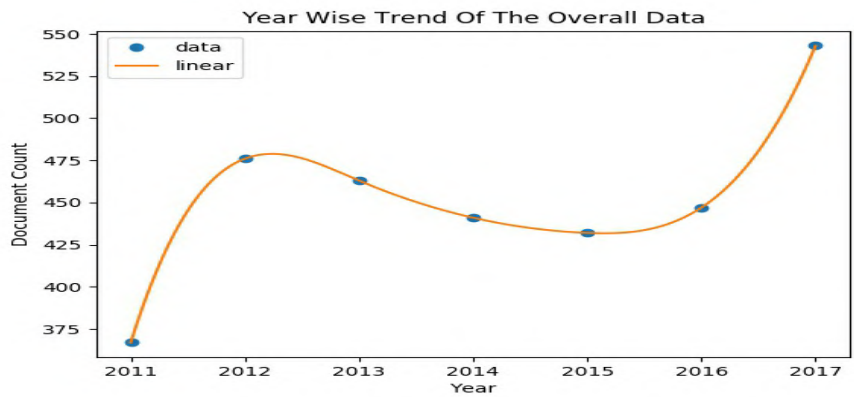

Figure 1.3 Year Wise trend of the Overall Data from 2011-2017

# APPENDIX – B

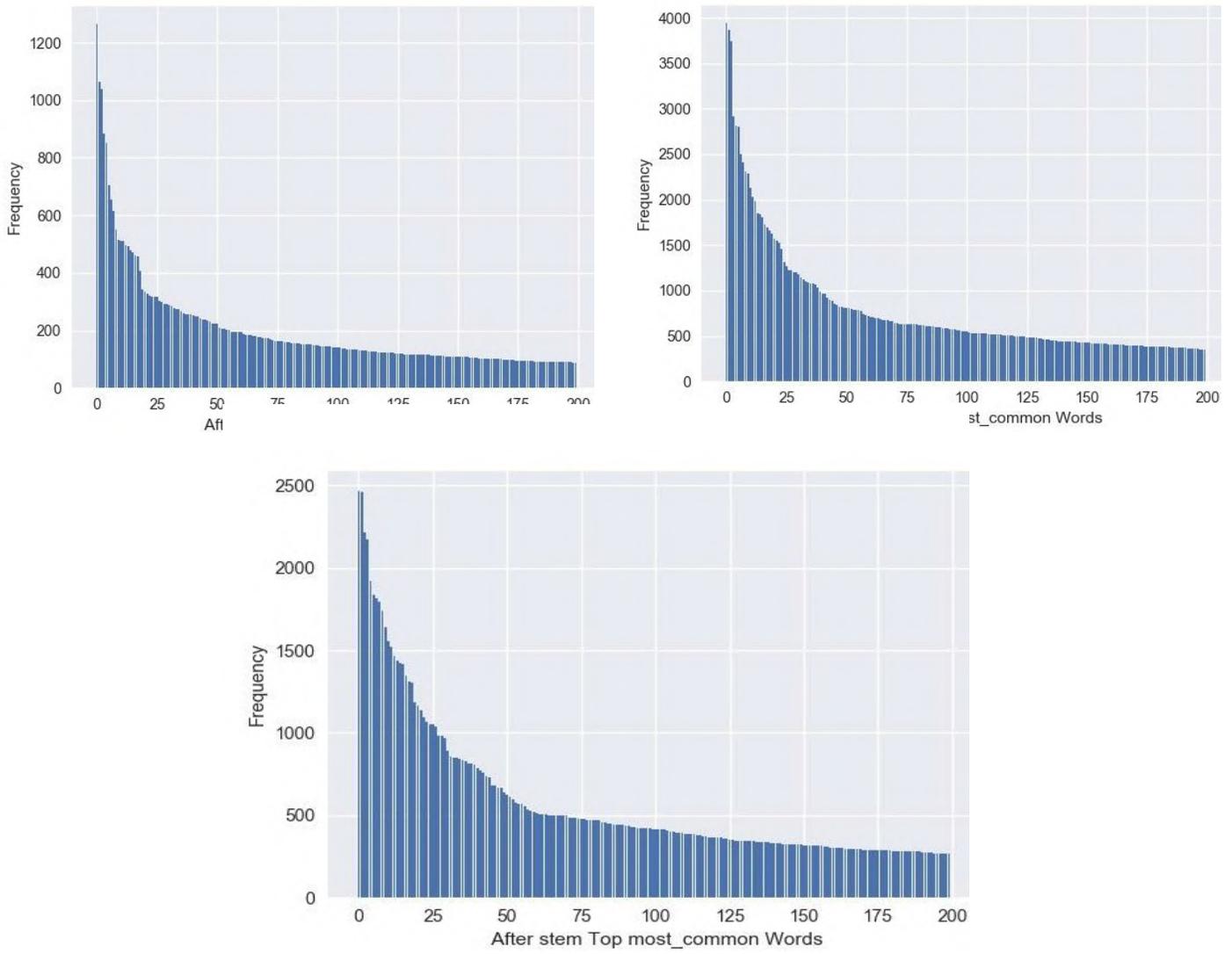

Figure 2.1 Frequency Distribution of top 200 Words Over 3 year Slots

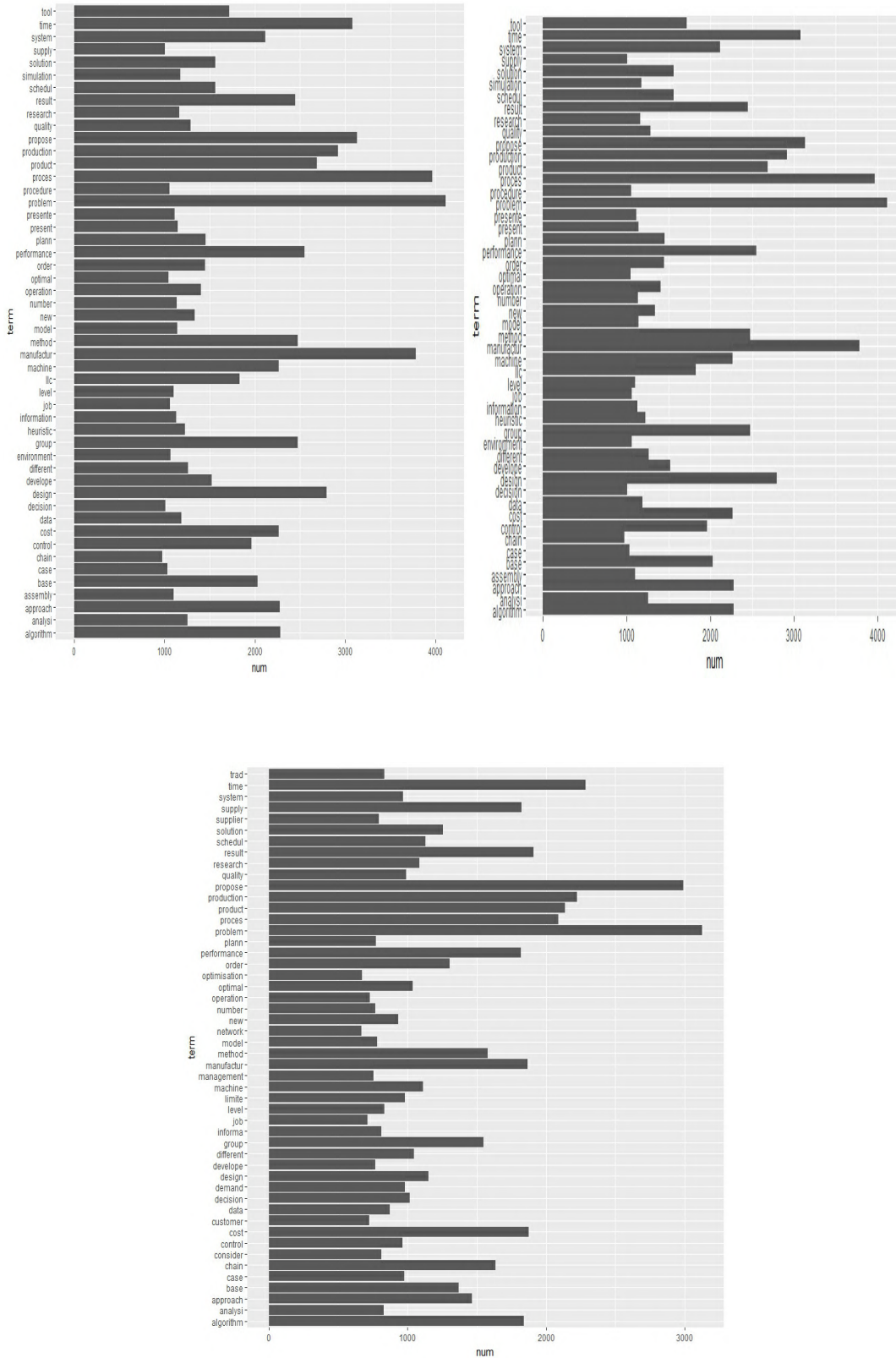

Figure 2.2 Term Frequency Bar Chart for R Programming from 3 Major Year Slots

# APPENDIX – C

Figure 3.1 Word Cloud for Different Year Slots by Python Program

Figure 3.2 Word Cloud for Different Year Slots by R Program

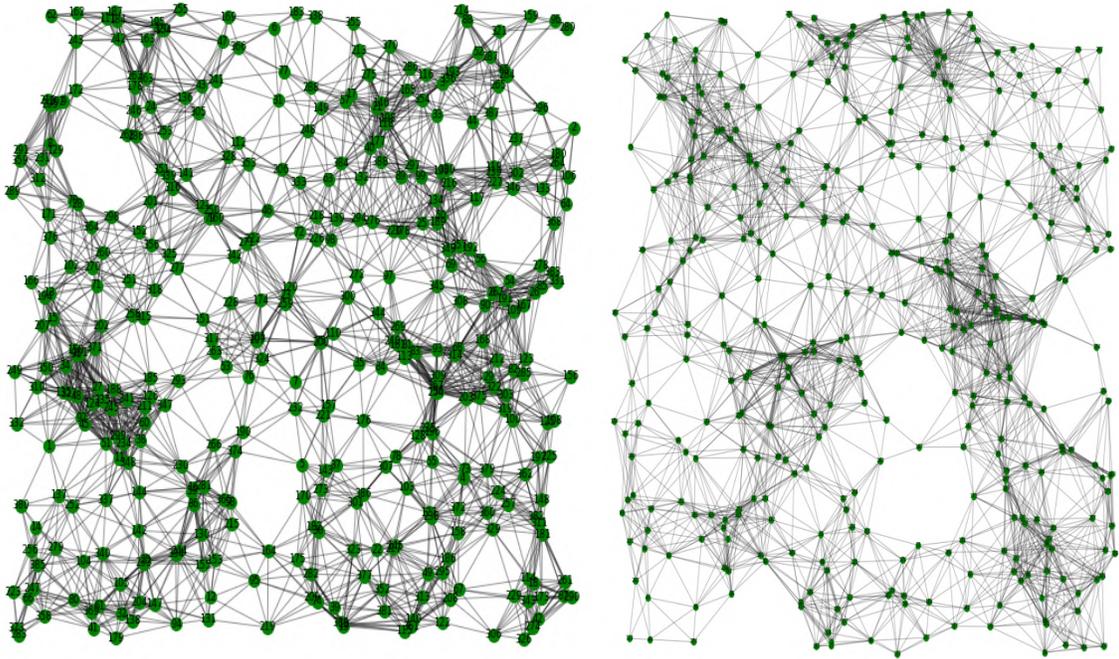

Figure 4.1 Overall Author Association from 1961-1990

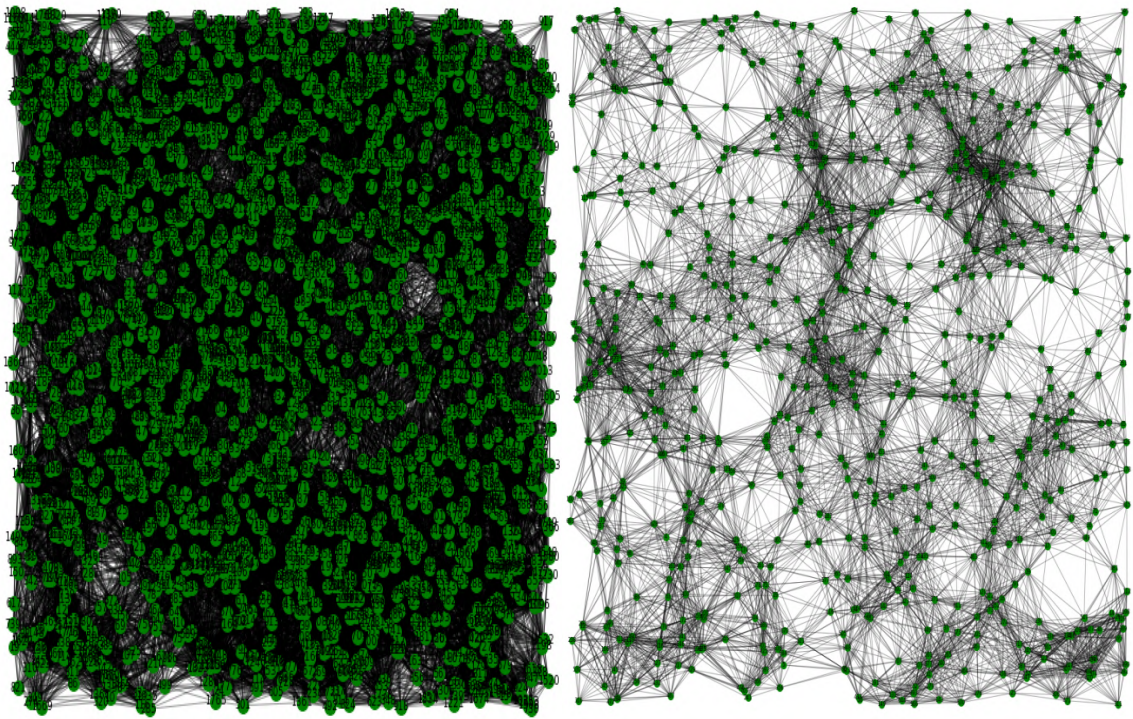

Figure 4.2 Overall Author Association from 1991-2010

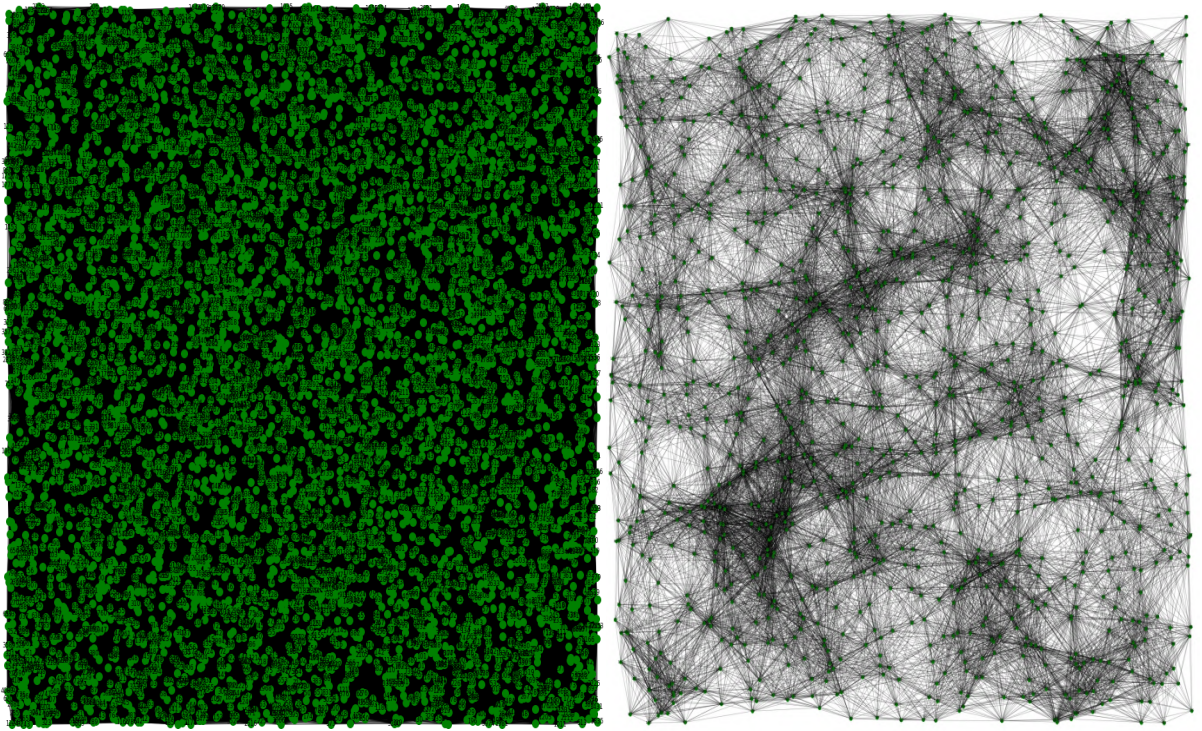

Figure 4.3 Overall Author Association from 2011-2017

Figure 5.1 Country Association for the Slot 1961-1990

Figure 5.2 Country Association for the Slot 1991-2010

Figure 5.3 Country Association For the slot 2011-2017

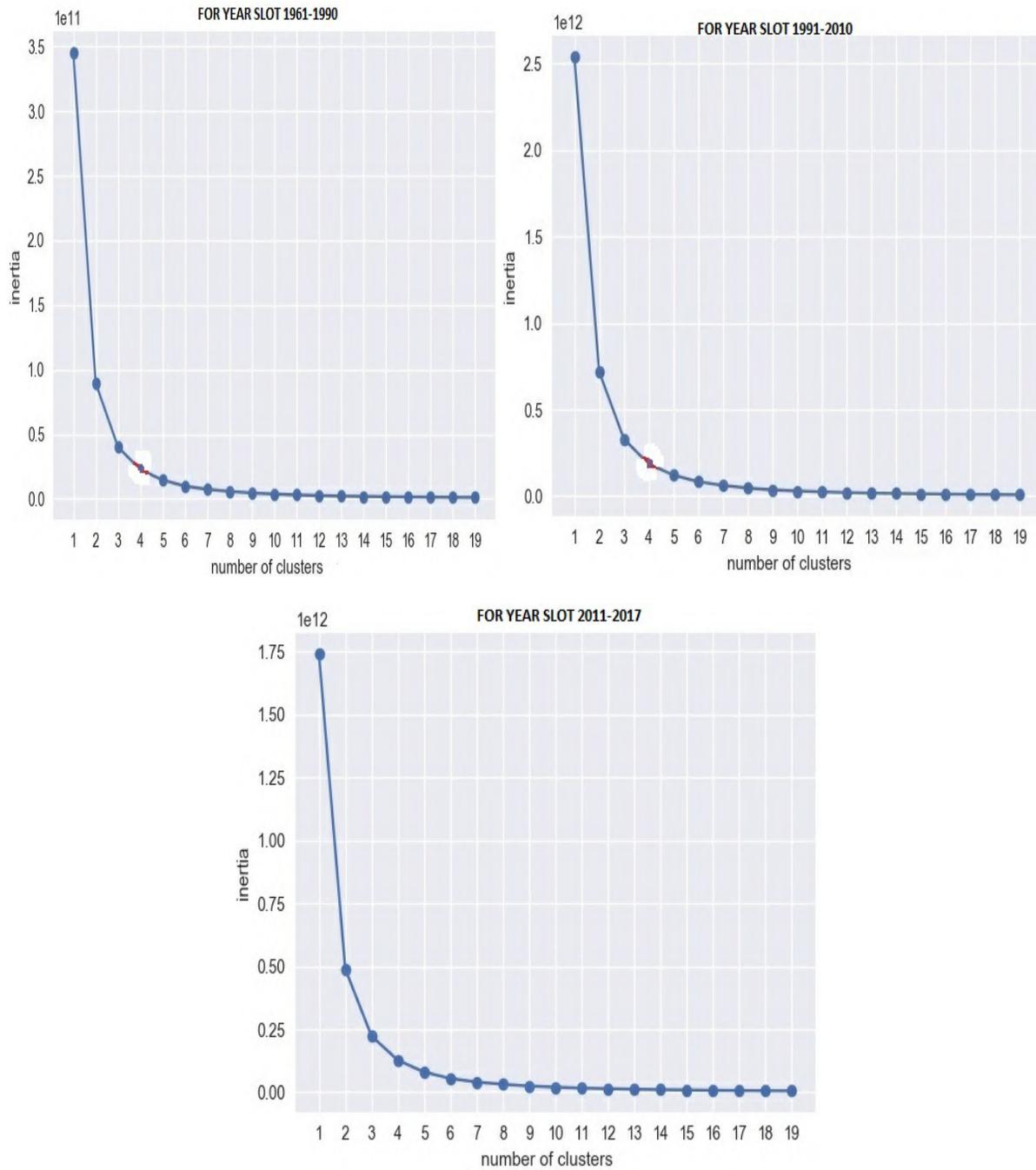

Figure 6.1 Number of clusters (k) and Inertia relationship by Python Program

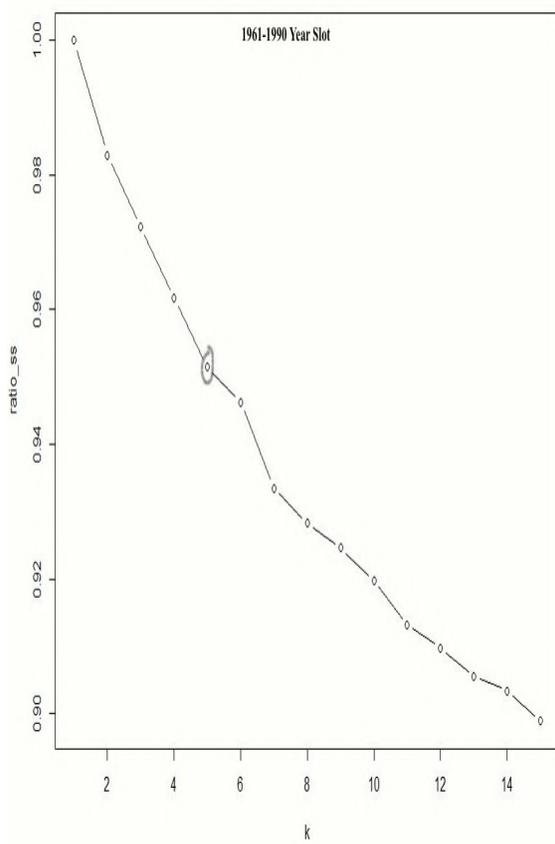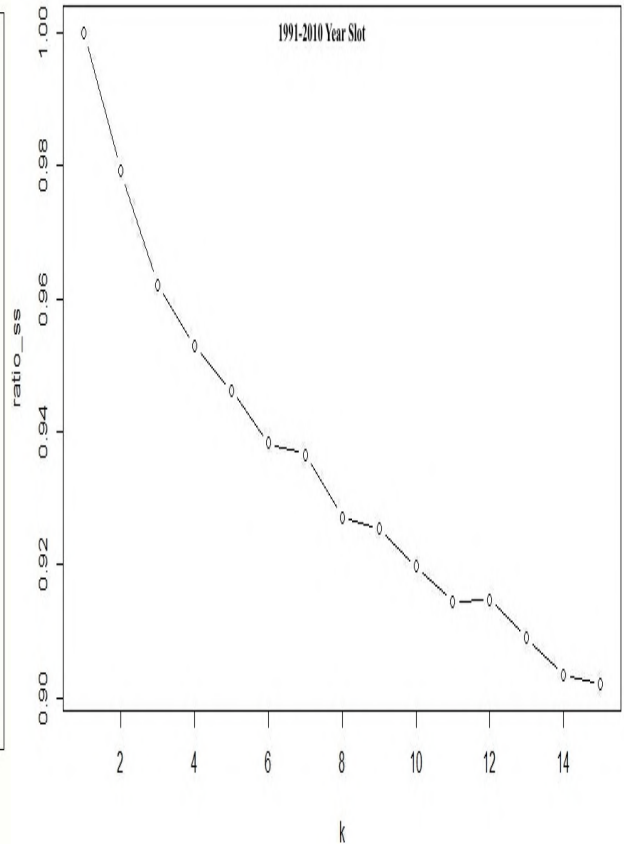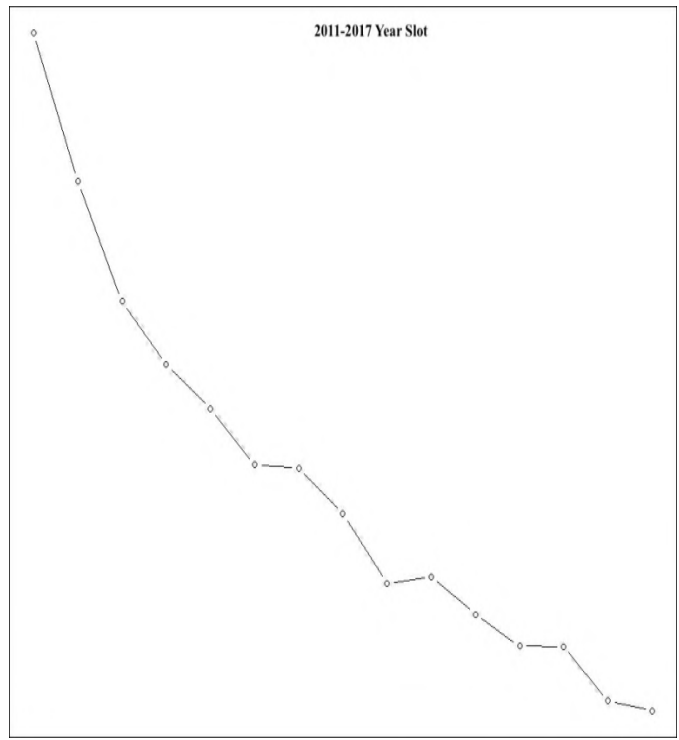

Figure 6.2 Number of clusters (k) and Inertia relationship by R Program

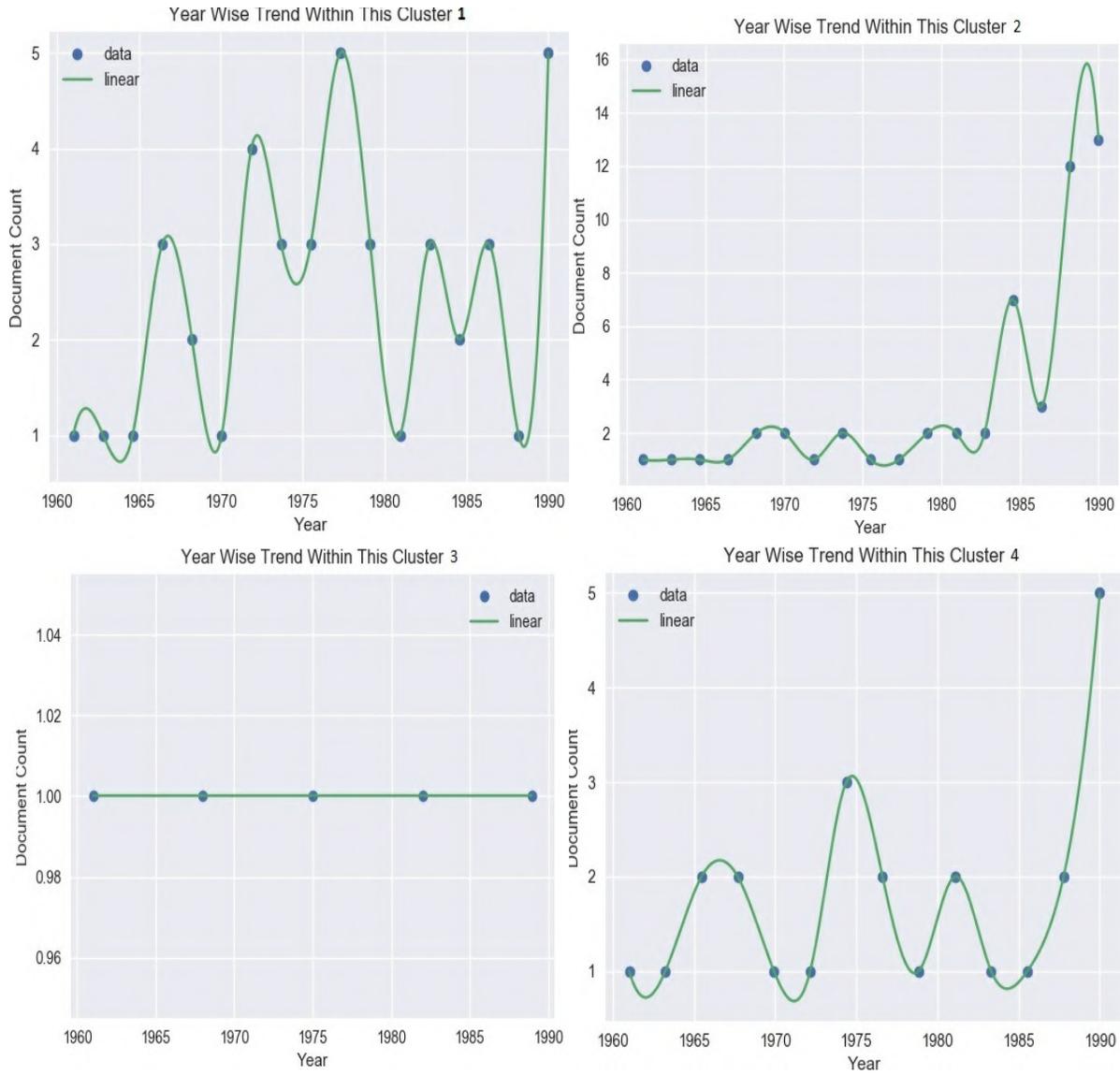

Figure 7.1 Clustering Trends by Python from 1961-1990

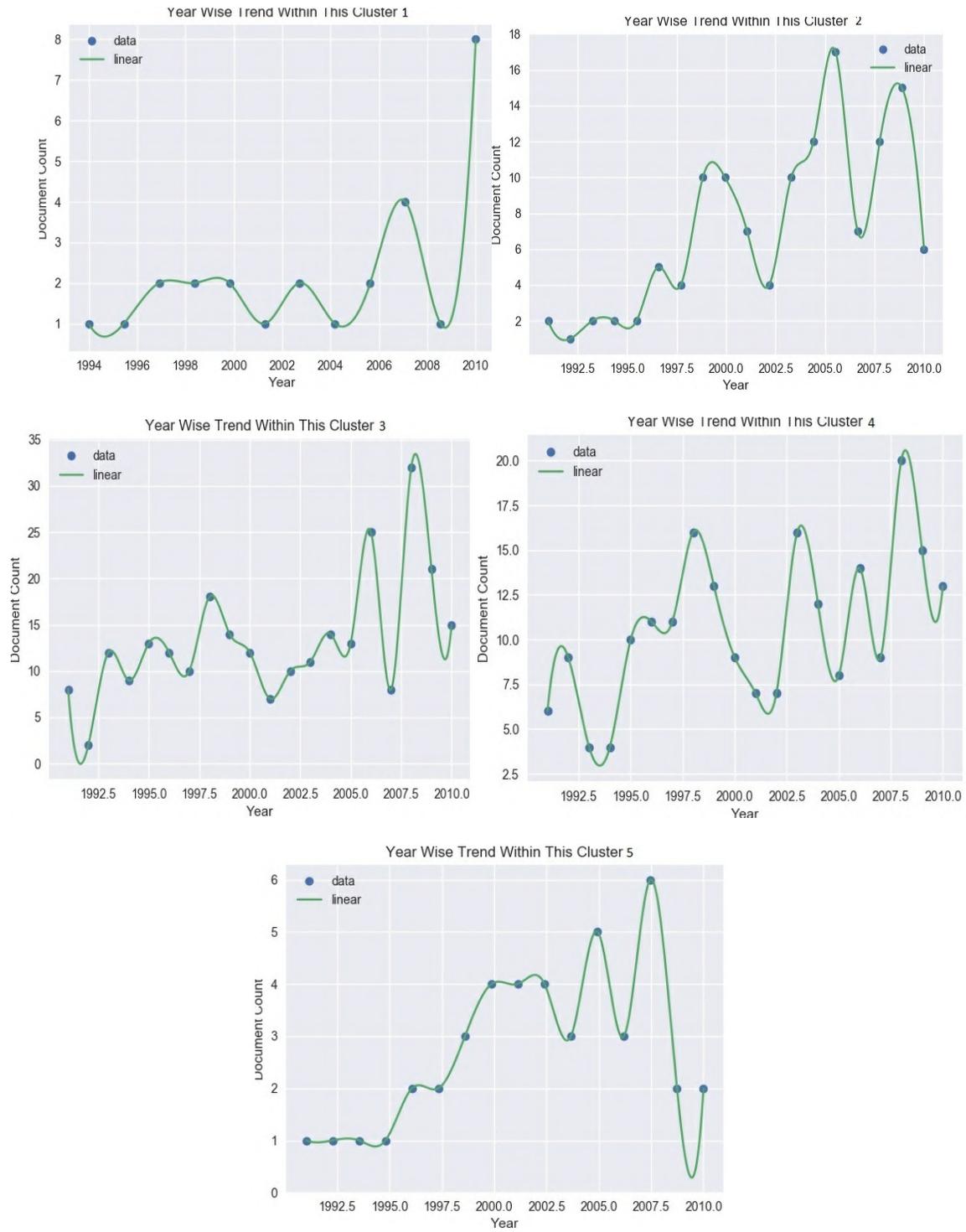

Figure 7.2 Clustering Trends by Python from 1991-2010

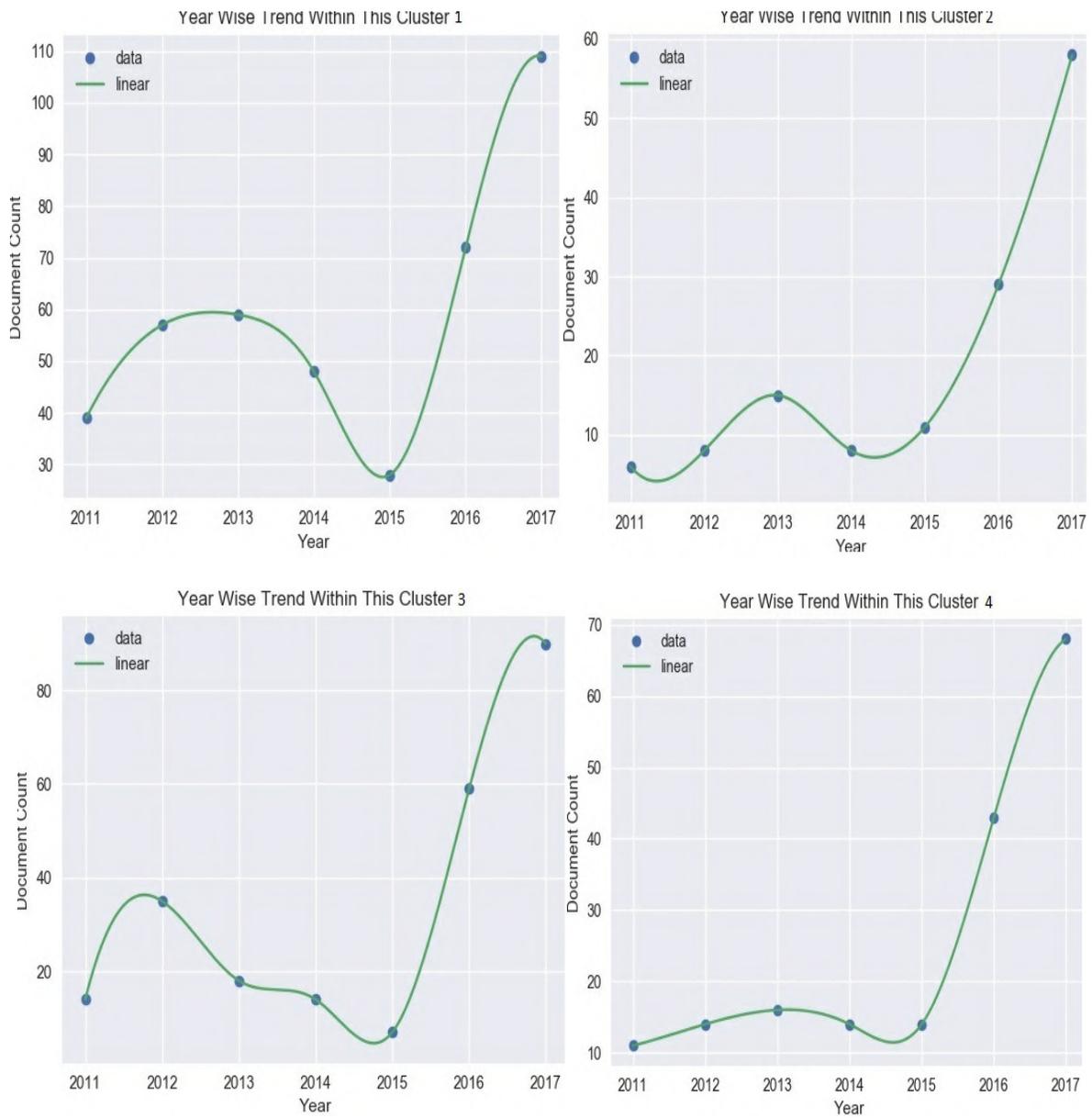

Figure 7.3 Clustering Trends by Python from 2011-2017

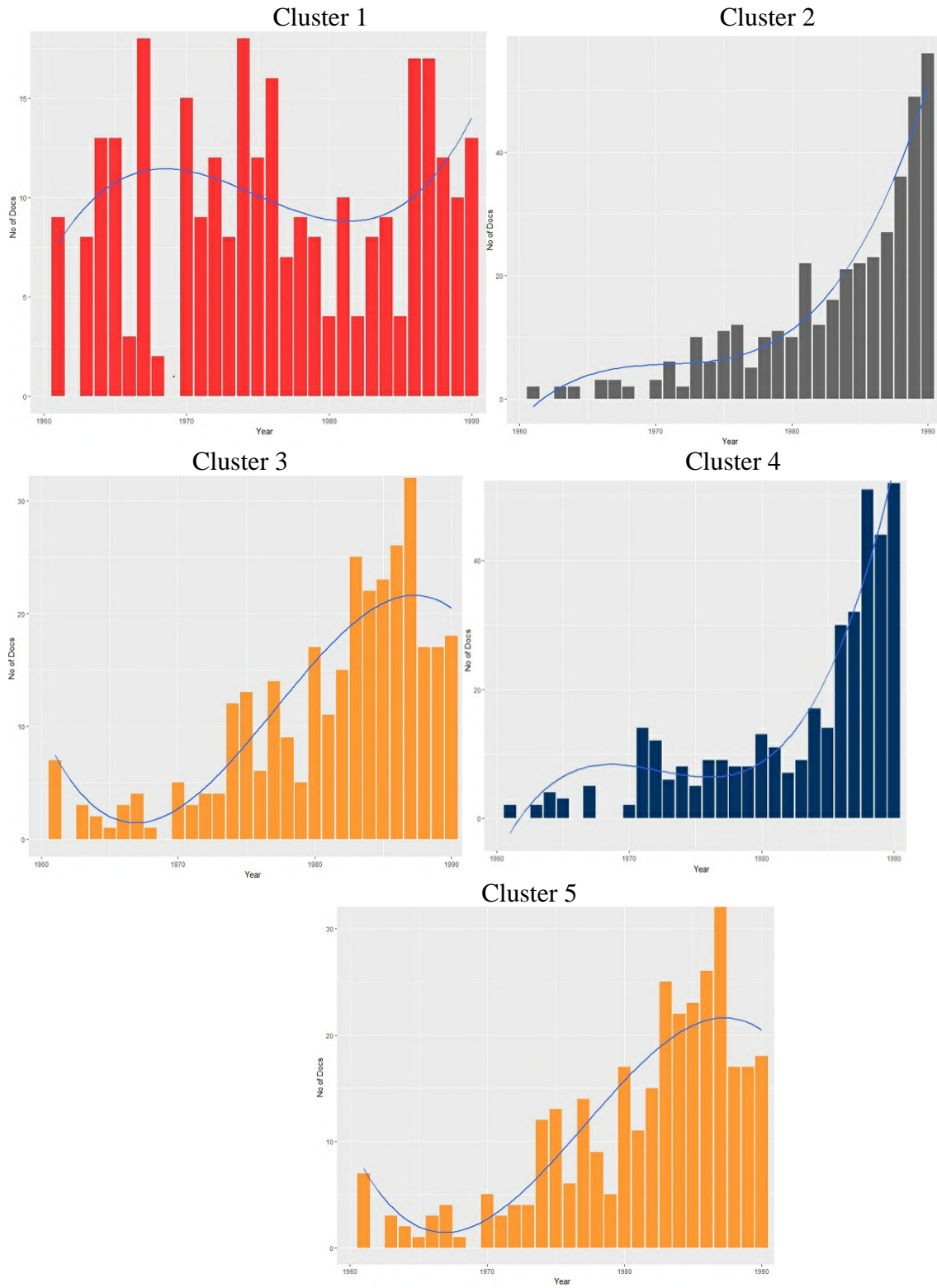

Figure 7.4 Clustering Trends by R from 1961-1990

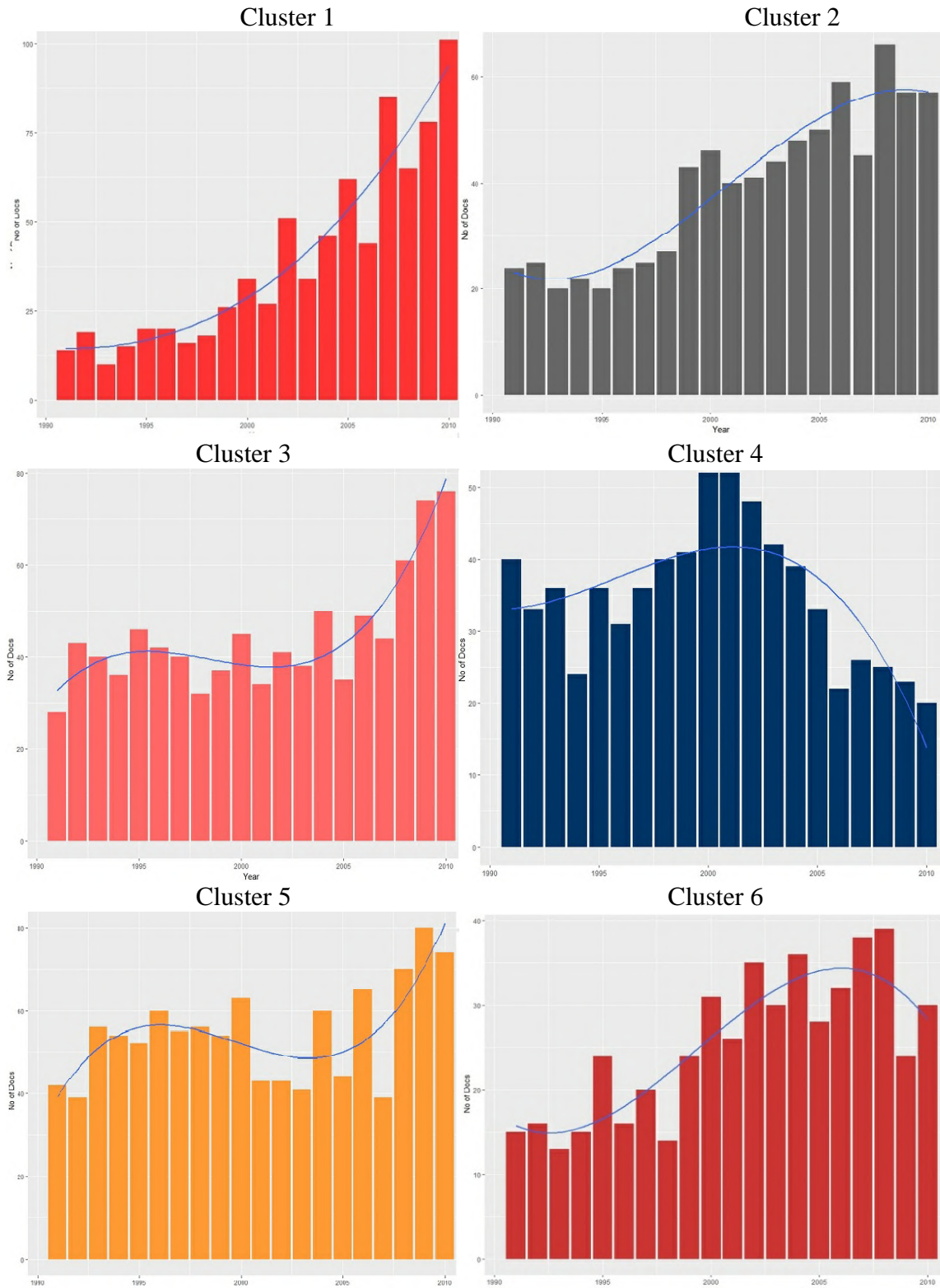

Figure 7.5 Clustering Trends by R from 1991-2010

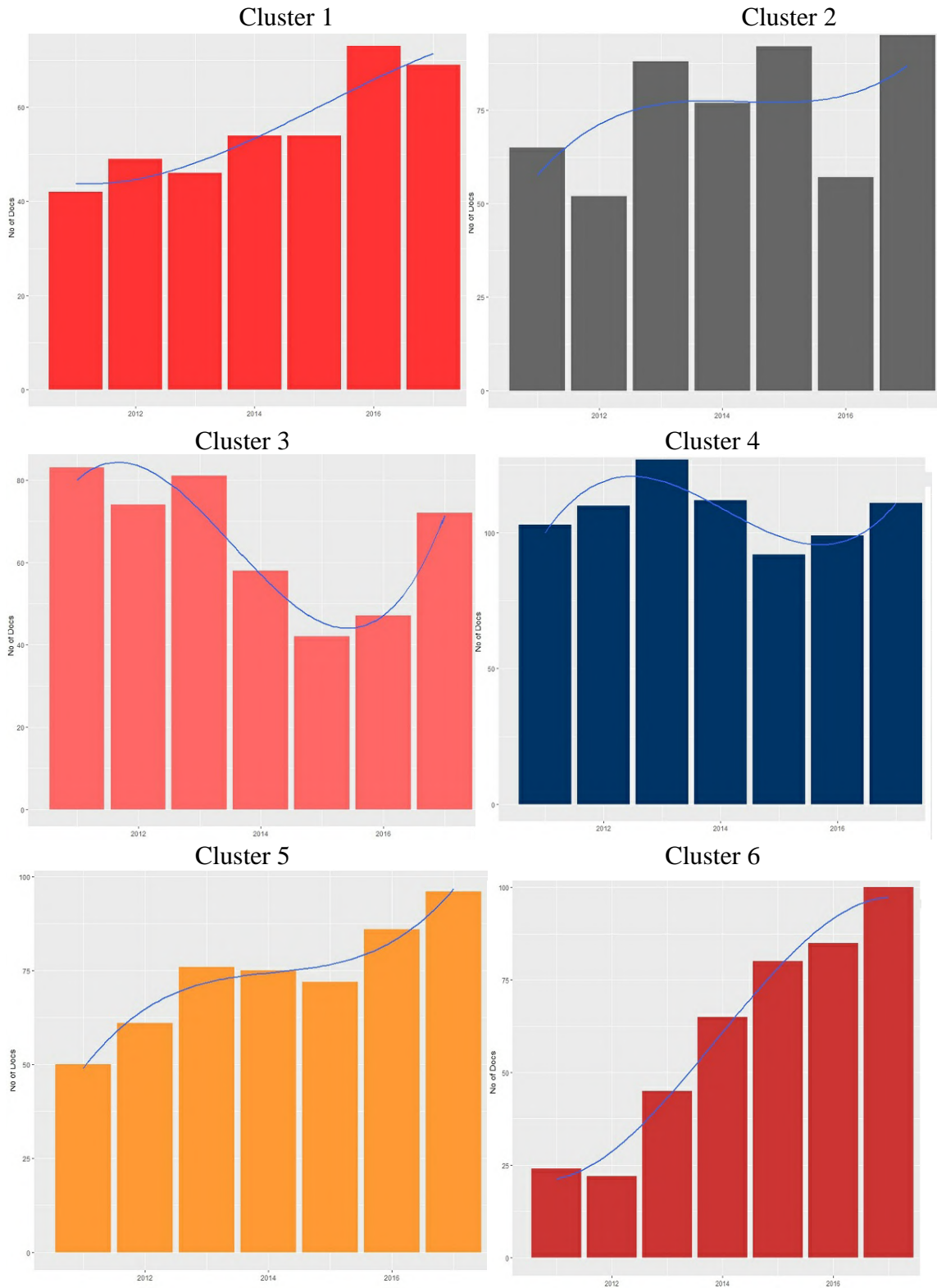

Figure 7.6 Clustering Trends by R from 2011-2017

1961-1990 Year Slot

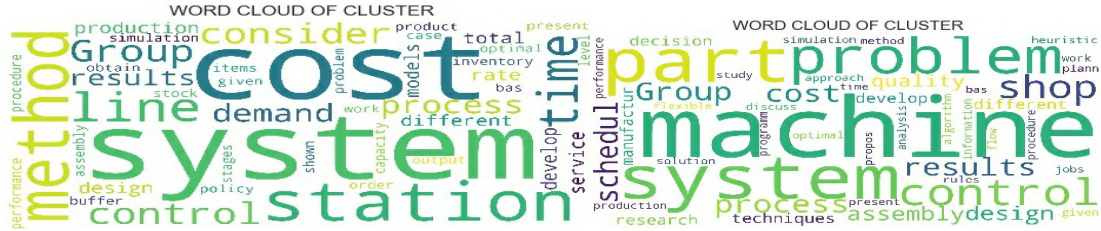

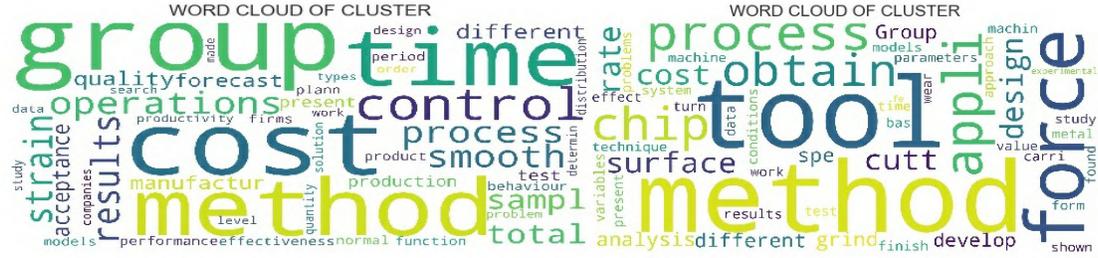

1991-2010 Year Slot

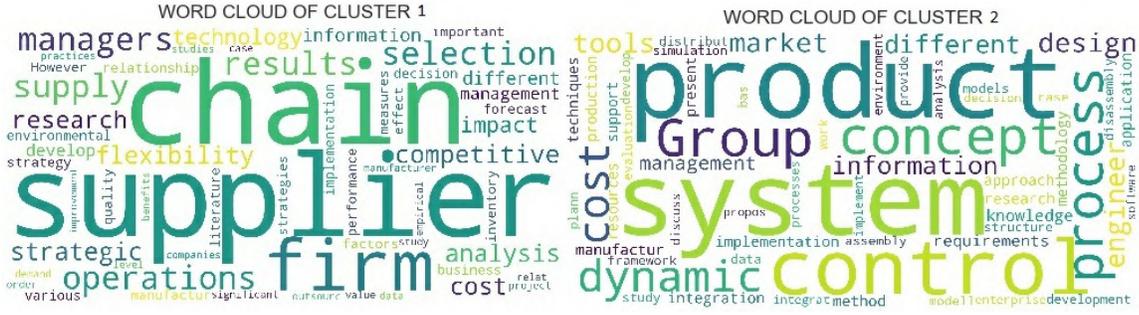

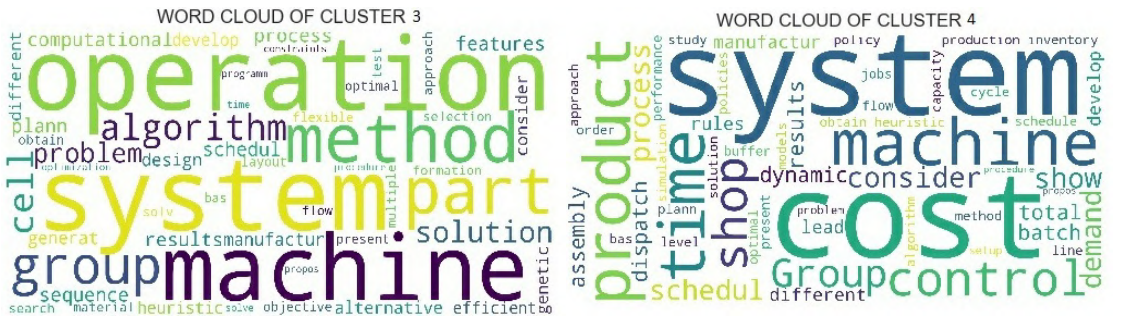

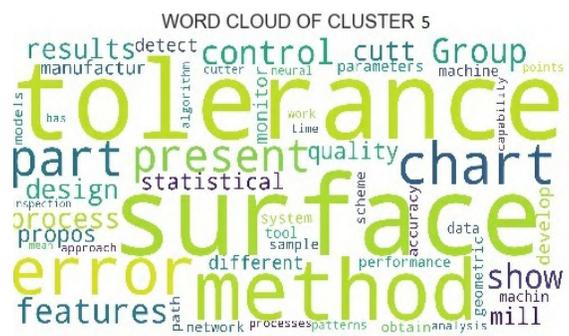

2011-2017 Year Slot

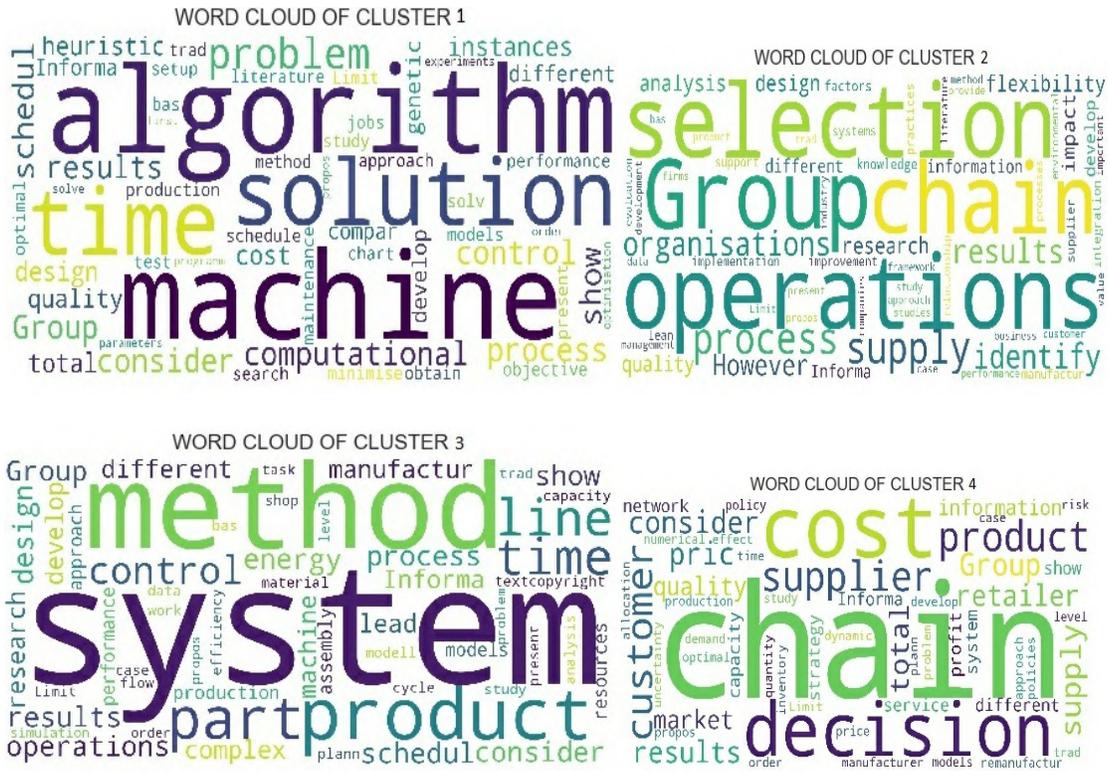

Figure 7.7 Word Clouds for Different Clusters by Python in 3 Year Slots

1961-1990 Year Slot

1991-2010 Year Slot

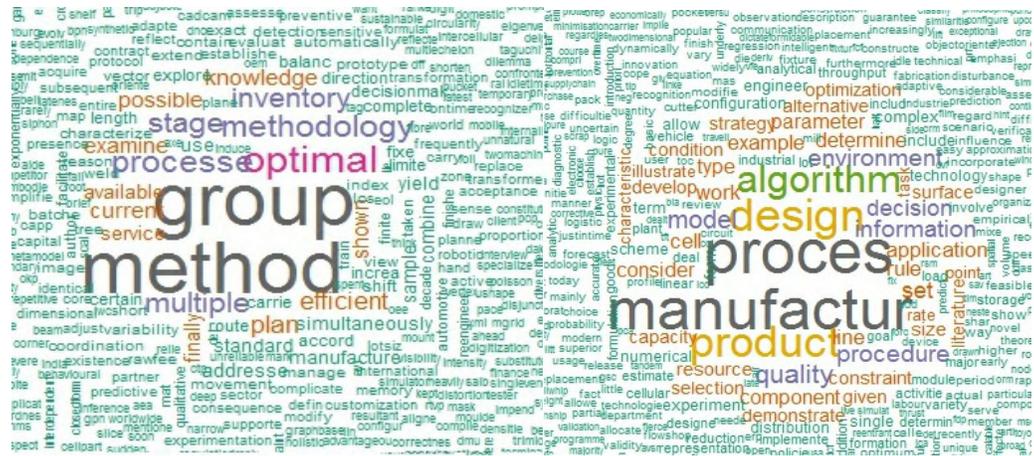

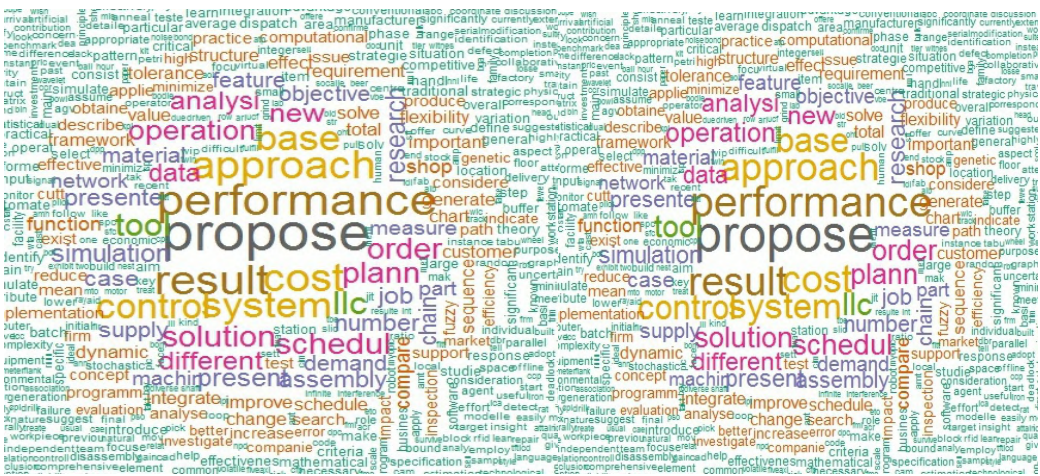

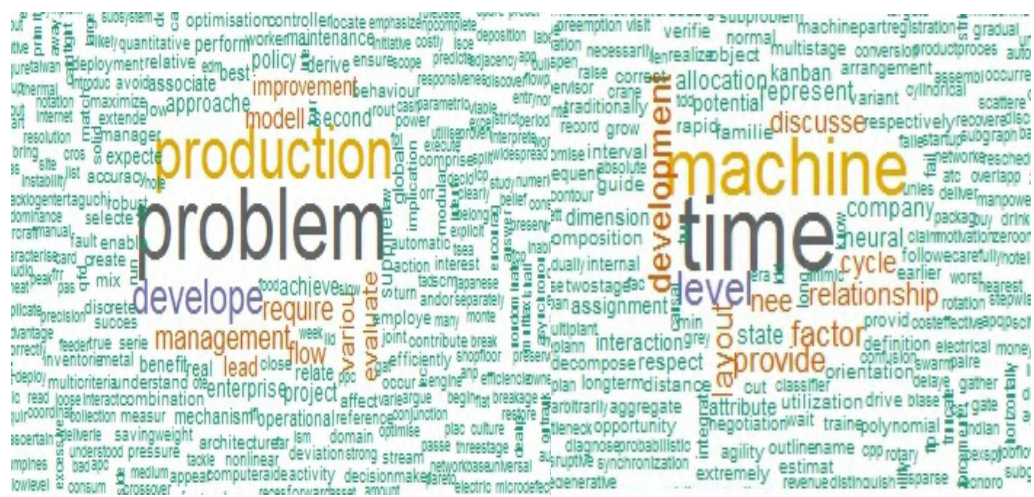

Figure 7.8 Word Clouds for Different Clusters by R in 3-Year Slots